\documentclass[journal,comsoc]{IEEEtran}
\usepackage{amsmath}
\usepackage{cite}
\usepackage{amssymb}
\usepackage{amsfonts}
\usepackage{algorithm}
\usepackage{dsfont}
\usepackage{graphicx}
\usepackage{epsfig}
\usepackage{subfigure}
\usepackage{psfrag}
\usepackage{xcolor}
\usepackage{url}
\usepackage[colorlinks,linkcolor=black,urlcolor=black,anchorcolor=black,citecolor=black,hyperfootnotes=true]{hyperref}

\title{{Channel Estimation for Intelligent Reflecting Surface Assisted Multiuser Communications: Framework, Algorithms, and Analysis}
\thanks{Manuscript received December 24, 2019, revised April 13, 2020, accepted June 15, 2020. The materials in this paper have been
presented in part at the IEEE Wireless Communications and Networking Conference (WCNC) 2020 \cite{Liu19}. This work was supported by the Hong Kong Polytechnic University under Research Grant P0030001. The work was also supported in part by the Key Area R\&D Program of Guangdong Province with grant No. 2018B030338001, by the National Key R\&D Program of China with grant No. 2018YFB1800800, by Natural Science Foundation of China with grant NSFC-61629101, and by Guangdong Zhujiang Project No. 2017ZT07X152. The associate editor coordinating the review of this paper and approving it for publication was Itsik Bergel. (Corresponding author: Liang Liu.)}
\thanks{Z. Wang and L. Liu are with the Department of Electronic and Information Engineering, the Hong Kong Polytechnic University, Hong Kong, China (e-mails: \{zhaorui.wang,liang-eie.liu\}@polyu.edu.hk).}
\thanks{S. Cui is with the  Shenzhen Research Institute of Big Data and the School of Science and Engineering, the Chinese University of Hong Kong, Shenzhen, China, 518172 (e-mail: shuguangcui@cuhk.edu.cn).}}
\author{\IEEEauthorblockN{Zhaorui Wang, Liang Liu, \IEEEmembership{Member, IEEE}, and Shuguang Cui, \IEEEmembership{Fellow, IEEE}}}

\begin{document}
\maketitle \thispagestyle{empty} \vspace{-0.3in}

\newtheorem{example}{Example}
\newtheorem{corollary}{Corollary}
\newtheorem{definition}{Definition}
\newtheorem{lemma}{Lemma}
\newtheorem{theorem}{Theorem}
\newtheorem{proposition}{Proposition}
\newtheorem{remark}{Remark}
\newcommand{\mv}[1]{\mbox{\boldmath{$ #1 $}}}

\begin{abstract}
In intelligent reflecting surface (IRS) assisted communication systems, the acquisition of channel state information is a crucial impediment for achieving the beamforming gain of IRS because of the considerable overhead required for channel estimation. Specifically, under the current beamforming design for IRS-assisted communications, in total $KMN+KM$ channel coefficients should be estimated, where $K$, $N$ and $M$ denote the numbers of users, IRS reflecting elements, and antennas at the base station (BS), respectively. For the first time in the literature, this paper points out that despite the vast number of channel coefficients that should be estimated, significant redundancy exists in the user-IRS-BS reflected channels of different users arising from the fact that each IRS element reflects the signals from all the users to the BS via the same channel. To utilize this redundancy for reducing the channel estimation time, we propose a novel three-phase pilot-based channel estimation framework for IRS-assisted uplink multiuser communications, in which the user-BS direct channels and the user-IRS-BS reflected channels of a typical user are estimated in Phase I and Phase II, respectively, while the user-IRS-BS reflected channels of the other users are estimated with low overhead in Phase III via leveraging their strong correlation with those of the typical user. Under this framework, we analytically prove that a time duration consisting of $K+N+\max(K-1,\lceil (K-1)N/M \rceil)$ pilot symbols is sufficient for perfectly recovering all the $KMN+KM$ channel coefficients under the case without receiver noise at the BS. Further, under the case with receiver noise, the user pilot sequences, IRS reflecting coefficients, and BS linear minimum mean-squared error channel estimators are characterized in closed-form.
\end{abstract}


\section{Introduction}\label{sec:Introduction}
\subsection{Motivation}

\begin{figure}[t]
  \centering
  \includegraphics[width=9cm]{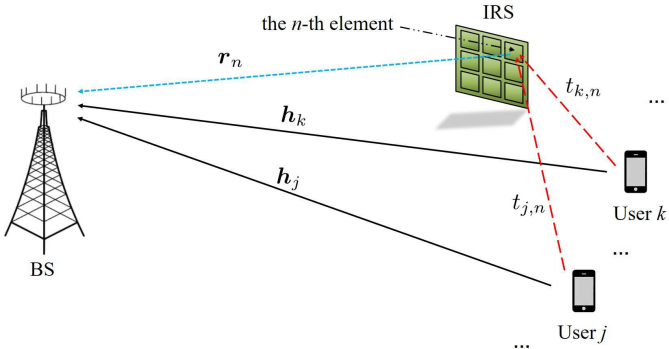}
  \caption{An IRS-assisted multiuser communication system: for any user $k$ and user $j$, the $n$th element of the IRS reflects their signals to the BS via the same channel.}\label{Fig1}
\end{figure}

Recently, intelligent reflecting surface (IRS) and its various equivalents have emerged as a promising solution to enhance the network throughput \cite{Liaskos08,Renzo19,Basar19,Larsson20}, thanks to their capability of modifying the wireless channels between the base station (BS) and users to be more favorable to communications via inducing phase shift to the incident signal at each reflecting element in real-time, as illustrated in Fig. \ref{Fig1}. Specifically, if perfect channel state information (CSI) is available, with the aid of a smart controller, the IRS is able to properly adjust its reflection coefficients at different reflecting elements based on the CSI such that the desired signals and interfering signals are added constructively and destructively at the receivers, respectively. Along this line, the IRS reflection coefficients optimization with perfect CSI has been widely studied under various setups (see, e.g., \cite{Wu18,Zhang19,Shuowen20,Schober19,Huang19,Guo,Hou,Wu19,Tang19,Fu19,Yu19}), where the effectiveness of IRS in enhancing the system throughput was verified.

However, the above throughput gain in IRS-assisted communication systems is critically dependent on the availability of CSI, the acquisition of which is quite challenging in practice. Particularly, to reduce the implementation cost, the IRS is generally not equipped with any radio frequency (RF) chains and thus not capable of performing any baseband processing functionality. Therefore, the user-IRS and IRS-BS channels cannot be separately estimated via traditional training-based approaches in general. Instead, only the concatenated user-IRS-BS channels can be estimated based on the training signals sent from the users/BS, and the corresponding number of channel coefficients can be quite large in practice. Specifically, consider a single cell consisting of a BS with $M$ antennas, $K$ single-antenna users, and one IRS with $N$ reflecting elements. It can be shown that the number of channel coefficients involved for designing the IRS reflection coefficients based on the algorithms proposed in \cite{Wu18,Zhang19} is $KMN+KM$. Considering the current paradigm shift towards massive multiple-input multiple-output (MIMO) \cite{Larsson14,Bjornson17,Bjornson16,Bjornson14}, i.e., a large $M$, and massive connectivity \cite{Liu18,Liu18Mag}, i.e., a large $K$, it is expected that the estimation of these channel coefficients can require tremendous time. This motivates us to devote our endeavour to developing efficient channel estimation strategies for IRS-assisted multiuser communication in this paper.

\subsection{Prior Work}

Recently, several works have proposed various strategies to efficiently estimate the channels for IRS-assisted communication \cite{Mishra19,Yang19,Zheng19,Jensen,Tellambura,Yuan19,Liu19,Liang19}. For the single-user system, in \cite{Mishra19,Yang19}, an on-off state control based channel estimation strategy was proposed, where only one IRS element is switched on at each time slot such that its reflected channel for the user can be estimated without interference from the reflected signals of the other IRS elements. Under this strategy, $N$ time slots are sufficient to perfectly estimate all the reflected channels for the user for the case without receiver noise at the BS. On the other hand, \cite{Zheng19,Jensen} proposed a novel discrete Fourier transform (DFT) based channel estimation strategy, in which all the IRS elements are on at each time slot, and their reflection coefficients are determined by the DFT matrix. Under this strategy, although still $N$ time slots are required to perfectly estimate all the reflected channels in the case without receiver noise at the BS, the channel estimation mean-squared error (MSE) in the case with receiver noise is significantly reduced compared to the on-off state control based strategy, because the IRS elements are always on to reflect all the signal power to the BS. Further, \cite{Tellambura} proposed a Lagrange-based estimation strategy to minimize the MSE for channel estimation. Last, \cite{Yuan19} formulated the single-user channel estimation problem as a combined sparse matrix factorization and matrix completion problem and proposed to apply the compressed sensing technique to solve the problem. On the other hand, for the multi-user case, our conference paper \cite{Liu19} pointed out the possibility of reducing the channel estimation time by utilizing the fact that the IRS reflects all users' signals to the BS via the same channels. However, there lacks a systematic study of this scheme. Moreover, \cite{Liang19} modeled the IRS reflected channels as sparse channels, and then applied the compressed sensing technique to estimate the channels with reduced time duration. However, whether the IRS reflected channels are sparse in practice is still unknown, while this paper aims to propose a channel estimation framework that can be applied to the more general channel model without assuming any channel property.

\subsection{Main Contributions}

In this paper, we consider an IRS-assisted multiuser uplink communication system where multiple single-antenna users communicate with a multi-antenna BS with the help of an IRS. Under this setup, we investigate the passive pilot based channel estimation approach, where the IRS elements passively reflect the pilot sequences sent by the users to the BS such that the BS is able to estimate the CSI associated with the IRS. The main contributions of this paper are summarized as follows.

First, we propose a novel \emph{three-phase channel estimation framework} for IRS-assisted multiuser uplink communications. The foundation of this framework lies in the \emph{correlation} among the user-IRS-BS reflected channels of different users: \emph{each IRS element reflects the signals from different users to the BS via the same channel}, as illustrated in Fig. \ref{Fig1}. To make the best use of this correlation, the proposed channel estimation framework works as follows. In Phase I, the IRS is switched off such that the BS can estimate its direct channels with the users. In Phase II, the IRS is switched on and only one typical user is selected to transmit non-zero pilot symbols such that its IRS reflected channels can be estimated. In Phase III, the other users transmit their pilot symbols and their IRS reflected channels can be efficiently estimated by exploiting the fact that these reflected channels are \emph{scaled versions} of the typical user's reflected channels and thus only the \emph{scaling factors (scalars)}, rather than the \emph{whole channels (vectors)}, need to be estimated.

Second, for the ideal case without receiver noise at the BS, we show that the theoretically minimal pilot sequence length to perfectly estimate all the channel coefficients under the proposed three-phase framework is $K+N+\max(K-1,\lceil(K-1)N/M \rceil)$. Specifically, it is shown that $K$ and $N$ time slots are sufficient to estimate the direct channels of all the users and IRS reflected channels of the typical user in Phase I and Phase II, respectively, while $\max(K-1,\lceil(K-1)N/M \rceil)$ time slots are sufficient for perfect channel estimation in Phase III. Interestingly, the minimum pilot sequence length decreases with $M$ generally. Such a result is in sharp contrast to the traditional multiuser channel estimation results without IRS, where the minimum pilot sequence length is independent of the number of receive antennas at the BS \cite{Hassibi03}.

Third, for the practical case with receiver noise at the BS, we design the linear minimum mean-squared error (LMMSE) channel estimation solutions in all the three phases. In each phase, the user transmit pilot, the IRS reflection coefficients, and the BS LMMSE channel estimators are characterized in closed-form. Moreover, the corresponding MSE for channel estimation is also derived to evaluate the performance of our proposed three-phase framework.

Last but not least, this paper reveals the significant role of \emph{massive MIMO} in channel estimation for IRS-assisted uplink communications. Especially, in the massive MIMO regime with $M>N$, the minimum pilot sequence length under our framework is $2K+N-1$, which is scalable with the number of users: if there is one more user in the system, only 2 additional pilot symbols are sufficient to estimate the new $MN+M$ channel coefficients associated with this user.

\subsection{Organization}
The rest of this paper is organized as follows.
Section \ref{sec:SYS} describes the system model for our considered IRS-assisted multiuser communication system.
Section \ref{sec:channel estimation model} introduces the three-phase channel estimation protocol.
Section \ref{sec:Performance Limits in Case without Noise} presents the minimum pilot sequence length for perfect channel estimation in closed-form for the case without noise at the BS.
Section \ref{sec:Channel Estimation in Case with Noise} proposes a closed-form solution of user pilot sequences, IRS reflecting coefficients and BS LMMSE channel estimators for the case with noise at the BS.
Numerical results are provided in Section \ref{sec:Numerical Examples}. Finally, Section \ref{sec:Conclusion} concludes
the paper and outlines the future research directions.

{\it Notation}: $\mv{I}$ and $\mv{0}$  denote an
identity matrix and an all-zero matrix, respectively, with
appropriate dimensions. For a square full-rank matrix $\mv{S}$, $\mv{S}^{-1}$
denotes its inverse. For a matrix
$\mv{M}$ of arbitrary size, $\mv{M}^{H}$, $\mv{M}^{T}$ and $\mv{M}^\ast$ denote the
conjugate transpose, transpose and conjugate of $\mv{M}$, respectively. Moreover, ${\rm
diag}(x_1,\cdots,x_K)$ denotes a diagonal matrix
with the diagonal elements given by $x_1,\cdots,x_K$. $\lceil x \rceil$ and $\lfloor x \rfloor$ denote the smallest integer that is no smaller than $x$ and the largest integer that is no larger than $x$, respectively.

\section{System Model}\label{sec:SYS}
We consider a narrow-band wireless system with $K$ single-antenna users simultaneously communicating with a BS in the uplink, where the BS is equipped with $M$ antennas. An IRS equipped with $N$ passive reflecting elements is deployed for enhancing the users' communication performance, as shown in Fig. \ref{Fig1}. We assume quasi-static block-fading channels, where all the channels remain approximately constant in each fading block. Moreover, the length of each fading block is denoted by $T$ symbols. Let $\mv{h}_k\in\mathbb{C}^{M\times1}$, $k=1,\cdots,K$, denote the direct channel from the $k$th user to the BS. Further, let $\mv{r}_n\in\mathbb{C}^{M\times1}$ and $t_{k,n}\in\mathbb{C}$ denote the channels from the $n$th IRS element to the BS and from the $k$th user to the $n$th IRS element, respectively, $k=1,\cdots,K$, $n=1,\dots,N$.

In this paper, $\mv{h}_k$'s are modeled as
\begin{align}\label{eqn:direct channel}
\mv{h}_k=(\mv{C}_k^{{\rm B}})^{\frac{1}{2}}\tilde{\mv{h}}_k, ~~~ \forall k,
\end{align}where $\mv{C}_k^{{\rm B}}\in \mathbb{C}^{M\times M}$ with $\mv{C}_k^{{\rm B}}\succ \mv{0}$ denotes the BS receive correlation matrix for user $k$, and $\tilde{\mv{h}}_k\sim \mathcal{CN}(\mv{0},\beta_k^{{\rm BU}}\mv{I})$ follows the independent and identically distributed (i.i.d.) Rayleigh fading channel model, with $\beta_k^{{\rm BU}}$ denoting the path loss of $\mv{h}_k$. Further, define $\mv{R}=[\mv{r}_1,\cdots,\mv{r}_N]$ as the overall channel from the IRS to the BS. Then, $\mv{R}$ is modeled by
\begin{align}\label{eqn:IRS to BS channel}
\mv{R}=(\mv{C}^{{\rm B}})^{\frac{1}{2}}\tilde{\mv{R}}(\mv{C}^{{\rm I}})^\frac{1}{2},
\end{align}where $\mv{C}^{{\rm B}}\in \mathbb{C}^{M\times M}$ with $\mv{C}^{{\rm B}}\succ \mv{0}$ and $\mv{C}^{{\rm I}}\in \mathbb{C}^{N\times N}$ with $\mv{C}^{{\rm I}}\succ \mv{0}$ denote the BS receive correlation matrix and IRS transmit correlation matrix for $\mv{R}$, respectively, and $\tilde{\mv{R}}\sim \mathcal{CN}(\mv{0},\beta^{{\rm BI}}N\mv{I})$ is the i.i.d. Rayleigh fading component, with $\beta^{{\rm BI}}$ denoting the path loss of $\mv{R}$. At last, define $\mv{t}_k=[t_{k,1},\cdots,t_{k,N}]^T$ as the overall channel from user $k$ to the IRS, $\forall k$. Then, $\mv{t}_k$'s are modeled as follows:
\begin{align}\label{eqn:user to IRS channel}
\mv{t}_k=(\mv{C}_k^{{\rm I}})^{\frac{1}{2}}\tilde{\mv{t}}_k, ~~~ \forall k,
\end{align}where $\mv{C}_k^{{\rm I}}\in \mathbb{C}^{N\times N}$ with $\mv{C}_k^{{\rm I}}\succ \mv{0}$ denotes the IRS receive correlation matrix for $\mv{t}_k$, and $\tilde{\mv{t}}_k\sim \mathcal{CN}(\mv{0},\beta_k^{{\rm IU}}\mv{I})$ denotes the i.i.d. Rayleigh fading component, with $\beta_k^{{\rm IU}}$ denoting the pass loss of $\mv{t}_k$.

Thanks to the IRS controller, each element on IRS is able to dynamically adjust its reflection coefficient for re-scattering the electromagnetic waves from the users to the BS such that the useful signal and harmful interference can be added at the BS in constructive and destructive manners, respectively \cite{Liaskos08,Renzo19,Basar19}. Specifically, let $\phi_{n,i}$ denote the reflection coefficient of the $n$th IRS element at the $i$th time instant over the considered coherence block, $n=1,\cdots,N$, $i=1,\cdots,T$, which satisfies
\begin{equation}
|\phi_{n,i}|=\left\{\begin{array} {ll} 1, ~ {\rm if ~ element} ~ n ~ {\rm is ~ on ~ at ~ time ~ instant} ~ i, \\ 0, ~ {\rm otherwise}. \end{array} \right. \label{eq:Sys-1}
\end{equation}Thus, if an IRS element is on, it can only change the phase of its incident signal \cite{Wu18,Zhang19}.

With the existence of the IRS, the received signal of the BS at time instant $i$, $i=1,\cdots,T$, is the superposition of the signals from the users' direct communication links and the reflected links via the IRS, which is expressed as
\begin{align}
\mv{y}^{(i)}&=\sum_{k=1}^{K}\mv{h}_k\sqrt{p}x_k^{(i)}+\sum_{k=1}^{K}\sum_{n=1}^{N}\phi_{n,i}t_{k,n}\mv{r}_n\sqrt{p}x_k^{(i)}+\mv{z}^{(i)} \nonumber \\
&=\sum_{k=1}^{K}\left(\mv{h}_k+\sum_{n=1}^{N}\phi_{n,i}\mv{g}_{k,n}\right)\sqrt{p}x_k^{(i)}+\mv{z}^{(i)},
\label{eq:Sys-1.5}
\end{align}where $x_k^{(i)}$ and $\mv{z}^{(i)}\sim\mathcal{CN}\left(\mv{0},\sigma^2\mv{I}\right)$ denote the transmit signal of user $k$ and additive white Gaussian noise (AWGN) at the BS at time instant $i$, respectively, $p$ denotes the identical transmit power of the users, and\vspace{-6pt}
\begin{align}\label{eqn:effective channel}
\mv{g}_{k,n}=t_{k,n}\mv{r}_n, ~~~ \forall n,k, \vspace{-4pt}
\end{align}denotes the user-IRS-BS reflected channel from the $k$th user to the BS via the $n$th IRS element.

In this paper, we consider the legacy two-stage transmission protocol for the uplink communications, in which each coherence block of length $T$ symbols is divided into the channel estimation stage consisting of $\tau<T$ symbols and data transmission stage consisting of $T-\tau$ symbols. Specifically, in the channel estimation stage, each user $k$ is assigned with a pilot sequence consisting of $\tau$ symbols:
\begin{align}\label{eqn:pilot channel}\vspace{-6pt}
\mv{a}_k=[a_{k,1},\cdots,a_{k,\tau}]^T, ~~~ k=1,\cdots,K, \vspace{-6pt}
\end{align}where the norm of $a_{k,i}$ is either zero or one, $\forall k,i$. At time instant $i\leq \tau$, user $k$ transmits $x_{k,i}=a_{k,i}$ to the BS, and the received signal at the BS is
\begin{align}\label{eqn:received pilot}
\mv{y}^{(i)}=\sum_{k=1}^{K}\left(\mv{h}_k+\sum_{n=1}^{N}\phi_{n,i}\mv{g}_{k,n}\right)\sqrt{p}a_{k,i}&+\mv{z}^{(i)}, i\leq \tau.
\end{align}Define $\mv{Y}=[\mv{y}^{(1)},\cdots,\mv{y}^{(\tau)}]\in \mathbb{C}^{M\times \tau}$ as the overall received signal at the BS across all the $\tau$ time instants of the channel estimation stage. We should properly design user pilot symbols $a_{k,i}$'s and IRS reflection coefficients $\phi_{n,i}$'s such that the BS is able to estimate the CSI that is useful for the data transmission stage (will be introduced later) based on the received signal $\mv{Y}$ as well as its knowledge of $a_{k,i}$'s and $\phi_{n,i}$'s.

\begin{figure*}[t]
  \centering
  \includegraphics[width=16.5cm]{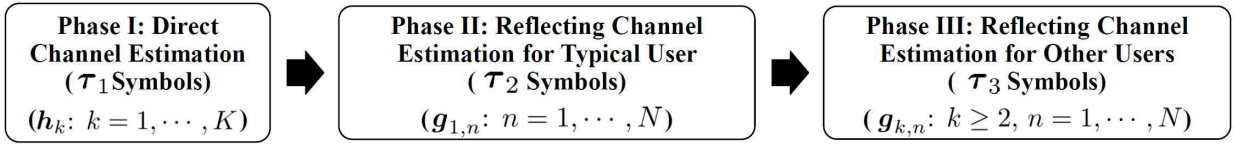}\vspace{-5pt}
  \caption{A flow graph for the proposed three-phase channel estimation framework.}\label{Fig2}
  \hrulefill \vspace{-10pt}
\end{figure*}

In the data transmission stage, the reflection coefficient of each IRS element $n$ is fixed over different time instants, i.e., $\phi_{n,i}=\phi_n$, $i=\tau+1,\cdots,T$ \cite{Wu18,Zhang19}. Moreover, to convey the information, the transmit symbol of user $k$ in the $i$th time instant is modeled as a circularly symmetric complex Gaussian (CSCG) random variable with zero mean and unit variance, i.e., $x_k^{(i)}\sim \mathcal{CN}(0,1)$, $k=1,\cdots,K$, $i=\tau+1,\cdots,T$. Then, at each time instant $i=\tau+1,\cdots,T$, the BS applies the beamforming vector $\mv{w}_k$ to decode the message of user $k$, i.e.,
\begin{align}
\mv{\tilde{y}}_k^{(i)}=\sum_{j=1}^K\mv{w}_k^H\left(\mv{h}_j\hspace{-2pt}+\hspace{-2pt}\sum_{n=1}^{N}\phi_n\mv{g}_{j,n}\right)\sqrt{p}x_j^{(i)}\hspace{-2pt}+\hspace{-2pt}\mv{w}_k^H\mv{z}^{(i)}.
\end{align}Therefore, the achievable rate of user $k$, $k=1,\cdots,K$, is
\begin{align}
R_k=\frac{T-\tau}{T}\log_2(1+\gamma_k),
	\label{eq:Sys-1.6}
\end{align}where $\frac{T-\tau}{T}$ denotes the fraction of time for data transmission, and
\begin{align}\label{eqn:gamma}
\gamma_k=\frac{p\left|\mv{w}_k^H\left(\mv{h}_k\hspace{-2pt}+\hspace{-2pt}\sum\limits_{n=1}^{N}\phi_n\mv{g}_{k,n}\right)\right|^2}{\sum\limits_{j\ne k}p\left|\mv{w}_k^H\left(\mv{h}_j\hspace{-2pt}+\hspace{-2pt}\sum\limits_{n=1}^{N}\phi_n\mv{g}_{j,n}\right)\right|^2\hspace{-2pt}+\hspace{-2pt}\sigma^2\|\mv{w}_k\|^2}.
\end{align}

It is observed from (\ref{eq:Sys-1.6}) and (\ref{eqn:gamma}) that to improve the user rate by jointly optimizing the receive beamforming vectors $\mv{w}_k$'s at the BS and reflection coefficients $\phi_n$'s at the IRS, in the channel training stage, the BS has to estimate $MK+MKN$ channel coefficients in $\mv{h}_k$'s, $k=1,\cdots,K$, and $\mv{g}_{k,n}$'s, $k=1,\cdots,K$, $n=1,\cdots,N$. Note that $MK+MKN$ is generally a very large number in next-generation cellular networks with large-scale antenna arrays at the BSs (i.e., a large $M$) and a massive number of connecting users (i.e., a large $K$). Further, to increase the time duration for data transmission in (\ref{eq:Sys-1.6}), very few pilot symbols can be utilized for estimating such a large number of channel coefficients. To tackle the above challenge, in the rest of this paper, we mainly focus on the channel training stage in our considered system, and propose an innovative scheme to estimate $\mv{h}_k$'s and $\mv{g}_{k,n}$'s accurately with low training overhead. The robust beamforming design for data transmission given the estimated channels with errors will be left to our future work.

\section{Three-Phase Channel Estimation Protocol}\label{sec:channel estimation model}
In this section, we propose a novel three-phase channel estimation protocol for IRS-assisted multiuser communications. The main idea is that although $KMN$ unknowns need to be estimated in $\mv{g}_{k,n}$'s, the degrees-of-freedom (DoF) for all these channel coefficients is much smaller than $KMN$. Specifically, each $\mv{r}_n$ appears in all $\mv{g}_{k,n}$'s, $k=1,\cdots,K$, according to (\ref{eqn:effective channel}), since each IRS element $n$ reflects the signals from all the $K$ users to the BS via the same channel $\mv{r}_n$. It is thus theoretically feasible to employ much fewer pilot symbols to estimate the $KMN$ correlated unknowns in $\mv{g}_{k,n}$'s. Nevertheless, it is challenging to practically exploit the correlations among the channel coefficients arising from $\mv{r}_n$'s, since the IRS cannot estimate $\mv{r}_n$'s due to the lack of RF chains.

In the following, we propose a novel three-phase channel estimation protocol, as depicted in Fig. \ref{Fig2}, which can exploit the channel correlations arising from $\mv{r}_n$'s to reduce the estimation time even without knowing $\mv{r}_n$'s. Specifically, in Phase I consisting of $\tau_1$ symbols, define
\begin{align}
\mv{a}_k^{{\rm I}}=[a_{k,1},\cdots,a_{k,\tau_1}]^T, ~~~ k=1,\cdots,K,
\end{align}as the pilot sequence sent by user $k$. In this phase, all the IRS elements are switched off, i.e.,
\begin{align}\label{eqn:off}
\phi_{n,i}=0, ~~~ n=1,\cdots,N, ~ i=1,\cdots,\tau_1.
\end{align}Then, the received signal at the BS at time slot $i$ of Phase I is
\begin{align}\label{eqn:direct channel received signal}
\mv{y}^{(i)}=\sum\limits_{k=1}^K\mv{h}_k\sqrt{p}a_{k,i}+\mv{z}^{(i)}, ~~~ i=1,\cdots,\tau_1.
\end{align}The BS thus needs to estimate the direct channels $\mv{h}_k$'s based on the above received signals.

In Phase II consisting of $\tau_2$ symbols, define
\begin{align}
\mv{a}_k^{{\rm II}}=[a_{k,\tau_1+1},\cdots,a_{k,\tau_1+\tau_2}]^T, ~~~ k=1,\cdots,K,
\end{align}as the pilot sequence of user $k$. In this phase, all the IRS reflection elements are switched on, and merely one \emph{typical user}, denoted by user $1$ for convenience, transmits non-zero pilot symbols to the BS, i.e.,
\begin{align}\label{eqn:zero pilot phase 2}
\mv{a}_k^{{\rm II}}=\mv{0},  ~~~ k=2,\cdots,K.
\end{align}Then, the received signal at the BS in time slot $i$ ($i=\tau_1+1,\cdots,\tau_1+\tau_2$) of Phase II is
\begin{align}\label{eqn:reflection channel user 1 received signal}
\mv{y}^{(i)}=\sum_{n=1}^{N}\phi_{n,i}\mv{g}_{1,n}\sqrt{p}a_{1,i}+\mv{h}_1\sqrt{p}a_{1,i}+\mv{z}^{(i)}.
\end{align}Based on its received signals in Phase II as well as knowledge of $\mv{h}_1$ after Phase I, the BS estimates the IRS reflected channels of this typical user, i.e., $\mv{g}_{1,n}$'s, $\forall n$.

In Phase III consisting of $\tau_3=\tau-\tau_1-\tau_2$ symbols, define
\begin{align}
\mv{a}_k^{{\rm III}}=[a_{k,\tau_1+\tau_2+1},\cdots,a_{k,\tau_1+\tau_2+\tau_3}]^T, ~ k=1,\cdots,K,
\end{align}as the pilot sequence of user $k$. In this phase, merely user 2 to user $K$ transmit the non-zero pilot symbols to the BS, i.e.,
\begin{align}\label{eqn:zero pilot phase 3}
\mv{a}_1^{{\rm III}}=\mv{0}.
\end{align}As a result, the received signal at the BS in time slot $i$ ($i=\tau_1+\tau_2+1,\cdots,\tau_1+\tau_2+\tau_3$) of Phase III is
\begin{align}\label{eqn:reflection channel other user received signal}
\mv{y}^{(i)}=\sum\limits_{k=2}^K\sum_{n=1}^{N}\phi_{n,i}\mv{g}_{k,n}\sqrt{p}a_{k,i}+\sum\limits_{k=2}^K\mv{h}_k\sqrt{p}a_{k,i}+\mv{z}^{(i)}.
\end{align}Intuitively, there are $(K-1)MN$ unknowns to be estimated in $\mv{g}_{k,n}$'s, $k\geq 2$. However, this number can be significantly reduced based on the following relationship between the user-IRS-BS reflected channels of user 1 and the other users:
\begin{align}\label{eqn:correlated channel}
\mv{g}_{k,n}=\lambda_{k,n}\mv{g}_{1,n}, ~~~ k=2,\cdots,K, ~ n=1,\cdots,N,
\end{align}where
\begin{align}\label{eqn:alpha}
\lambda_{k,n}=\frac{t_{k,n}}{t_{1,n}}, ~~~ k=2,\cdots,K, ~ n=1,\cdots,N.
\end{align}By taking (\ref{eqn:correlated channel}) into (\ref{eqn:reflection channel other user received signal}), the received signal at the BS in the $i$th time slot of Phase III reduces to
\begin{align}\label{eqn:reflection channel other user received signal equivalent}
\hspace{-5pt}\mv{y}^{(i)}\hspace{-2pt}=\hspace{-2pt}\sum\limits_{k=2}^K\sum_{n=1}^{N}\hspace{-2pt}\phi_{n,i}\mv{g}_{1,n}\sqrt{p}a_{k,i}\lambda_{k,n}\hspace{-2pt}+\hspace{-2pt}\sum\limits_{k=2}^K\mv{h}_k\sqrt{p}a_{k,i}\hspace{-2pt}+\hspace{-2pt}\mv{z}^{(i)}.
\end{align}With known $\mv{g}_{1,n}$'s, $n=1,\cdots,N$, after Phase II, (\ref{eqn:reflection channel other user received signal equivalent}) reveals that each channel vector $\mv{g}_{k,n}$ with $M$ unknowns, $k\geq 2$, can be efficiently recovered via merely estimating a scalar $\lambda_{k,n}$. As a result, the number of unknowns to be estimated in Phase III is significantly reduced.

For the purpose of drawing essential insights, in the rest of this paper, we first introduce this protocol for the ideal case without noise at the BS, i.e., $\mv{z}^{(i)}=\mv{0}$, $\forall i$, and characterize the \emph{minimum pilot sequence length} to estimate all the channels perfectly. Such a result can theoretically demonstrate the performance gain brought by this new protocol for channel estimation in our considered system. Then, we will illustrate how to implement our proposed three-phase channel estimation framework in the practical case with noise at the BS and characterize the \emph{MSE} for estimating the channels.

\section{Performance Limits for Case without Noise}\label{sec:Performance Limits in Case without Noise}
In this section, we consider the ideal case without noise at the BS. In this scenario, the proposed three-phase channel estimation protocol works as follows.

\subsection{Phase I: Direct Channel Estimation}\label{sec1}\vspace{-2pt}
In the case without noise, i.e., $\mv{z}^{(i)}=0$, $\forall i$, according to \cite{Hassibi03}, each user $k$ can send
\begin{align}\label{eqn:length of phase 1}
\tau_1\geq \tilde{\tau}_1=K,
\end{align}pilot symbols to the BS for channel estimation. Then, based on (\ref{eqn:direct channel received signal}), the received signal at the BS over the whole phase is
\begin{align}\label{eqn:received pilot direct channel}
\mv{Y}^{{\rm I}}&=[\mv{y}^{(1)},\cdots,\mv{y}^{(\tau_1)}] \nonumber \\ &=\sqrt{p}[\mv{h}_1,\cdots,\mv{h}_K][(\mv{a}_1^{{\rm I}}),\cdots,(\mv{a}_K^{{\rm I}})]^T.
\end{align}As a result, the direct channels $\mv{h}_k$'s can be estimated perfectly by solving (\ref{eqn:received pilot direct channel}) if the pilot sequences of different users are orthogonal to each other, i.e.,
\begin{align}\label{eqn:full rank}
[(\mv{a}_1^{{\rm I}}),\cdots,(\mv{a}_K^{{\rm I}})]^T[(\mv{a}_1^{{\rm I}})^\ast,\cdots,(\mv{a}_K^{{\rm I}})^\ast]=\tau_1\mv{I}.
\end{align}Since each pilot sequence consists of $\tau_1\geq K$ symbols, it is feasible to design $\mv{a}_k^{{\rm I}}$'s to satisfy (\ref{eqn:full rank}). Then, according to (\ref{eqn:received pilot direct channel}), $\mv{h}_k$'s can be perfected estimated as
\begin{align}\label{eqn:perfect channel estimation phase 1}
[\mv{h}_1,\cdots,\mv{h}_K]=\frac{1}{\tau_1\sqrt{p}}\mv{Y}^{{\rm I}}[(\mv{a}_1^{{\rm I}})^\ast,\cdots,(\mv{a}_K^{{\rm I}})^\ast].
\end{align}

\subsection{Phase II: Reflecting Channel Estimation for Typical User}\label{sec2}\vspace{-2pt}
In the second phase, to estimate $\mv{g}_{1,n}$'s, we only allow the typical user, denoted by user 1, to transmit its pilot, as shown in (\ref{eqn:zero pilot phase 2}). Note that $\mv{h}_k$'s have already been perfectly estimated by (\ref{eqn:perfect channel estimation phase 1}) in Phase I. Therefore, their interference for estimating $\mv{g}_{1,n}$'s can be canceled from the received signal at the BS in Phase II. According to (\ref{eqn:reflection channel user 1 received signal}), after interference cancellation, the effective received signal at the BS at time instant $i$ ($i=\tau_1+1,\cdots,\tau_1+\tau_2$) in Phase II is
\begin{align}\label{eqn:received pilot reflected channel user 1}
\bar{\mv{y}}^{(i)}=\mv{y}^{(i)}-\sum_{k=1}^{K}\mv{h}_k\sqrt{p}a_{k,i} = \sum_{n=1}^{N}\phi_{n,i}\mv{g}_{1,n}\sqrt{p}a_{1,i}.
\end{align}The overall effective received signal at the BS in the second phase is then expressed as
\begin{align}\label{eqn:received pilot phase 2}
\bar{\mv{Y}}^{{\rm II}}&=[\bar{\mv{y}}^{(\tau_1+1)},\cdots,\bar{\mv{y}}^{(\tau_1+\tau_2)}] \nonumber \\
&= \sqrt{p}[\mv{g}_{1,1},\cdots,\mv{g}_{1,N}]\mv{\Phi}^{{\rm II}}{\rm diag}(\mv{a}_1^{{\rm II}}),
\end{align}where
\begin{align}\label{eqn:Phi}
\mv{\Phi}^{{\rm II}}=\left[\begin{array}{ccc}\phi_{1,\tau_1+1} & \cdots & \phi_{1,\tau_1+\tau_2} \\ \vdots & \ddots & \vdots \\ \phi_{N,\tau_1+1} & \cdots & \phi_{N,\tau_1+\tau_2} \end{array} \right].
\end{align}To solve (\ref{eqn:received pilot phase 2}), we can simply set
\begin{align}\label{eqn:pilot phase 2}
a_{1,i}=1, ~~~ i=\tau_1+1,\cdots,\tau_1+\tau_2.
\end{align}In this case, it can be shown that as long as
\begin{align}\label{eqn:tau 2}
\tau_2\geq \tilde{\tau}_2=N,
\end{align}we can always construct a $\mv{\Phi}^{{\rm II}}$ such that ${\rm rank}(\mv{\Phi}^{{\rm II}})=N$ under the constraint given in (\ref{eq:Sys-1}). The construction of such a $\mv{\Phi}^{{\rm II}}$ can be based on the DFT matrix
\begin{align}\label{eqn:DFT}
\hspace{-4pt} \mv{\Phi}^{{\rm II}}=\left[\begin{array}{ccccc} 1 & 1 & 1 & \cdots & 1 \\ 1 & \omega & \omega^2 & \cdots & \omega^{\tau_2-1} \\ 1 & \omega^2 & \omega^4 & \cdots & \omega^{2(\tau_2-1)} \\ \vdots & \vdots & \vdots & \ddots & \vdots \\ 1 & \omega^{N-1} & \omega^{2(N-1)} & \cdots & \omega^{(N-1)(\tau_2-1)} \end{array} \right],
\end{align}where $\omega=e^{-2\pi q/\tau_2}$ with $q^2=-1$. In this case, it follows that $\mv{\Phi}^{{\rm II}}(\mv{\Phi}^{{\rm II}})^H=\tau_2\mv{I}$. As a result, the reflected channels of user $1$ can be perfectly estimated as
\begin{align}\label{eqn:perfect channel estimation phase 2}
[\mv{g}_{1,1},\cdots,\mv{g}_{1,N}]=\frac{1}{\tau_2\sqrt{p}}\bar{\mv{Y}}^{{\rm II}}(\mv{\Phi}^{{\rm II}})^H.
\end{align}

\begin{figure*}[t]
\setcounter{equation}{36}
\begin{align}\label{eqn:pilot array}
& \hspace{-4pt}{\small \mv{V}}\hspace{-4pt}=\hspace{-4pt}\left[\begin{array}{ccccccc} \hspace{-4pt}\phi_{1,\theta+1}a_{2,\theta+1}{\small \mv{g}_{1,1}} \hspace{-4pt}&\hspace{-4pt} \cdots \hspace{-4pt}& \hspace{-4pt} \phi_{N,\theta+1}a_{2,\theta+1}{\small \mv{g}_{1,N}} \hspace{-4pt}&\hspace{-4pt} \cdots\hspace{-4pt} & \hspace{-4pt} \phi_{1,\theta+1}a_{K,\theta+1}{\small \mv{g}_{1,1}} \hspace{-4pt} & \hspace{-4pt} \cdots \hspace{-4pt} & \hspace{-4pt} \phi_{N,\theta+1}a_{K,\theta+1}{\small \mv{g}_{1,N}} \hspace{-4pt} \\ \hspace{-4pt}\vdots \hspace{-4pt}& \hspace{-4pt}\ddots \hspace{-4pt}&\hspace{-4pt} \vdots \hspace{-4pt}&\hspace{-4pt} \ddots \hspace{-4pt}&\hspace{-4pt} \vdots \hspace{-4pt}& \hspace{-4pt}\ddots \hspace{-4pt}&\hspace{-4pt} \vdots \hspace{-4pt} \\ \hspace{-4pt} \phi_{1,\theta+\tau_3}a_{2,\theta+\tau_3}{\small \mv{g}_{1,1}} \hspace{-4pt}&\hspace{-4pt} \cdots \hspace{-4pt}& \hspace{-4pt} \phi_{N,\theta+\tau_3}a_{2,\theta+\tau_3}{\small \mv{g}_{1,N}} \hspace{-4pt}& \hspace{-4pt}\cdots \hspace{-4pt} & \hspace{-4pt} \phi_{1,\theta+\tau_3}a_{K,\theta+\tau_3}{\small \mv{g}_{1,1}} \hspace{-4pt}&\hspace{-4pt} \cdots \hspace{-4pt}& \hspace{-4pt} \phi_{N,\theta+\tau_3}a_{K,\theta+\tau_3}{\small \mv{g}_{1,N}} \hspace{-4pt} \end{array} \right], \\ \nonumber \\
& \hspace{-4pt}{\rm with} ~ \theta=\tau_1+\tau_2. \nonumber
\end{align}
\hrulefill
\end{figure*}

\subsection{Phase III: Reflecting Channel Estimation for Other Users}
To estimate the channels in Phase III, one straightforward approach is to allow only one user $k\geq 2$ to transmit $\tau_2 \geq N$ pilot symbols each time such that its reflected channels $\mv{g}_{k,n}$'s, $\forall n$, can be directly estimated based on the approach for estimating $\mv{g}_{1,n}$'s. Under such a scheme, we need to use at least $\tau_3=(K-1)N$ time instants in total to estimate the reflecting channels of the remaining $K-1$ users. However, with a large number of users, the estimation of $\mv{g}_{k,n}$'s will take quite a long time, which leads to reduced user transmission rate due to the limited time left for data transmission as shown in (\ref{eq:Sys-1.6}).

We propose to exploit the channel correlations among $\mv{g}_{k,n}$'s to reduce the channel estimation time in Phase III. Specifically, similar to (\ref{eqn:received pilot reflected channel user 1}), after cancelling the interference caused by the direct channels $\mv{h}_k$'s from (\ref{eqn:reflection channel other user received signal equivalent}), the effective received signal at the BS at each time instant $i$, $i=\tau_1+\tau_2+1,\cdots,\tau_1+\tau_2+\tau_3$, in Phase III is
\begin{align}\setcounter{equation}{34}
\hspace{-5pt}\bar{\mv{y}}^{(i)}\hspace{-2pt}=\hspace{-2pt}\mv{y}^{(i)}\hspace{-2pt}-\hspace{-2pt}\sum\limits_{k=2}^{K}\mv{h}_k\sqrt{p}a_{k,i}\hspace{-2pt} = \hspace{-2pt} \sum\limits_{k=2}^K\sum\limits_{n=1}^{N}\hspace{-2pt}\phi_{n,i}\sqrt{p}a_{k,i}\lambda_{k,n}\mv{g}_{1,n}. \label{eqn:received pilot reflected channel user k 2}
\end{align}The overall effective received signal at the BS in Phase III is
\begin{align}\label{eqn:received pilot phase 3}
\hspace{-6pt}\bar{\mv{y}}^{{\rm III}}\hspace{-2pt}=\hspace{-2pt}\left[\left(\bar{\mv{y}}^{(\tau_1\hspace{-1pt}+\hspace{-1pt}\tau_2\hspace{-1pt}+\hspace{-1pt}1)}\right)^T,\cdots,\left(\bar{\mv{y}}^{(\tau_1\hspace{-1pt}+\hspace{-1pt}\tau_2\hspace{-1pt}+\hspace{-1pt}\tau_3)}\right)^T\right]^T\hspace{-2pt}=\hspace{-2pt}\sqrt{p}\mv{V}\mv{\lambda},
\end{align}where $\mv{\lambda}=[\mv{\lambda}_2^T,\cdots,\mv{\lambda}_K^T]^T\in \mathbb{C}^{(K-1)N\times 1}$ with $\mv{\lambda}_k=[\lambda_{k,1} \cdots \lambda_{k,N}]^T$, $k=2,\cdots,K$, and $\mv{V}\in \mathbb{C}^{M\tau_3\times (K-1)N}$ is given in (\ref{eqn:pilot array}) on the top of this page.

Mathematically, (\ref{eqn:received pilot phase 3}) defines an equivalent linear channel estimation model consisting of $(K-1)N$ users, where each column of $\mv{V}$ denotes the pilot sequence sent by each of these effective users. One interesting observation of $\mv{V}$ here is that thanks to the multiple antennas at the BS, the effective channel estimation time is increased from $\tau_3$ to $M\tau_3$. In other words, it is possible to leverage the multi-antenna technology to significantly reduce the channel training time in Phase III under our proposed strategy.

In the following, via a proper design of $\mv{a}_k^{{\rm III}}=[a_{k,\tau_1+\tau_2+1},\cdots,a_{k,\tau_1+\tau_2+\tau_3}]^T$'s, $k=2,\cdots,K$, and $\phi_{n,i}$'s, $n=1,\cdots,N$, $i=\tau_1+\tau_2+1,\cdots,\tau_1+\tau_2+\tau_3$, we aim to find the minimum value of $\tau_3$ to satisfy ${\rm rank}(\mv{V})=(K-1)N$ such that $\mv{\lambda}$ can be perfectly estimated based on (\ref{eqn:received pilot phase 3}). We start with the case of $M\geq N$.

\begin{theorem}\label{theorem1}
In the case of $M\geq N$, the minimum value of $\tau_3$ to guarantee perfect estimation of $\mv{\lambda}$ according to (\ref{eqn:received pilot phase 3}) is given by
\begin{align}\setcounter{equation}{37}\label{eqn:tau 3}
\tilde{\tau}_3=K-1.
\end{align}To achieve perfect estimation of $\mv{\lambda}$ given the above minimum value of $\tau_3$, in the case of $M\geq N$, we can set
\begin{align}
& \hspace{-2pt} a_{k,i}=\left\{\begin{array}{ll}1, & {\rm if} ~ k-1=i-\tau_1-\tau_2, \\ 0, & {\rm otherwise}, \end{array}\right. ~ 2\leq k \leq K, \label{eqn:a case 1}  \\
& \hspace{-2pt} \phi_{n,i}=1, ~ 1\leq n \leq N, ~ \tau_1\hspace{-2pt}+\hspace{-2pt}\tau_2\hspace{-2pt}+\hspace{-2pt}1\leq i \leq \tau_1\hspace{-2pt}+\hspace{-2pt}\tau_2\hspace{-2pt}+\hspace{-2pt}K\hspace{-2pt}-\hspace{-2pt}1. \label{eqn:b case 1}
\end{align}Then, $\mv{\lambda}$ can be perfectly estimated as
\begin{align}\label{eqn:estimation 1}
\hspace{-2pt}\mv{\lambda}_k=[\mv{g}_{1,1},\cdots,\mv{g}_{1,N}]^{\dag}\frac{\bar{\mv{y}}^{(\tau_1+\tau_2+k-1)}}{\sqrt{p}}, ~ k=2,\cdots,K,
\end{align}where for any matrix $\mv{B}\in \mathbb{C}^{x\times y}$ with $x\geq y$, $\mv{B}^\dag=(\mv{B}^H\mv{B})^{-1}\mv{B}^H$ denotes its pseudo-inverse matrix.
\end{theorem}

\begin{IEEEproof}
Please refer to Appendix \ref{appendix1}.
\end{IEEEproof}

Next, we consider the case of $M<N$. In this case, define $\rho=\lfloor\frac{N}{M}\rfloor$, $\upsilon=N-M\rho$, and $\mathcal{N}=\{1,\cdots,N\}$. For each user $k\geq 2$, define two sets $\Lambda_{k,1}\subset \mathcal{N}$ with cardinality $|\Lambda_{k,1}|=N-\upsilon$ and $\Lambda_{k,2}\subset \mathcal{N}$ with cardinality $|\Lambda_{k,2}|=\upsilon$, which are constructed as follows. First, define
\begin{align}\label{eqn:T}
\mathcal{T}_k=\{(k-2)\upsilon +1,\cdots,(k-1)\upsilon\}, ~~~ k=2,\cdots,K.
\end{align}Then, we construct $\Lambda_{k,1}$'s and $\Lambda_{k,2}$'s as
\begin{align}
& \Lambda_{k,2}=\{m-(\lceil \frac{m}{N} \rceil-1)N:\forall m\in \mathcal{T}_k\}, \label{eqn:Lambda 2} \\
& \Lambda_{k,1}=\mathcal{N}\setminus \Lambda_{k,2}, ~~~ k=2,\cdots,K. \label{eqn:Lambda 1}
\end{align}Moreover, for any $i=1,\cdots, (K-1)\rho$, define $\kappa_i=(i-(\lceil\frac{i}{\rho}\rceil-1)\rho-1)M$ and
\begin{align}\label{eqn:Omega}
\Omega_i=\{\Lambda_{\lceil \frac{i}{\rho}\rceil+1,1}(\kappa_i+1),\cdots,\Lambda_{\lceil \frac{i}{\rho}\rceil+1,1}(\kappa_i+M)\},
\end{align}where given any set $\mathcal{B}$, $\mathcal{B}(i)$ denotes its $i$th element. While for any $i=(K-1)\rho+1,(K-1)\rho+2,\cdots$, define
\begin{align}\label{eqn:J}
\mathcal{J}_i=&\{(i\hspace{-2pt}-\hspace{-2pt}(K\hspace{-2pt}-\hspace{-2pt}1)\rho\hspace{-2pt}-\hspace{-2pt}1)M\hspace{-2pt}+\hspace{-2pt}1,\cdots,\nonumber \\ & \min((i\hspace{-2pt}-\hspace{-2pt}(K\hspace{-2pt}-\hspace{-2pt}1)\rho)M,(K\hspace{-2pt}-\hspace{-2pt}1)N\hspace{-2pt}-\hspace{-2pt}(\hspace{-2pt}K-\hspace{-2pt}1)M\rho)\}.
\end{align}Based on $\mathcal{J}_i$, given any $i>(K-1)\rho$, we define
\begin{align}
& \mathcal{K}_i=\{\lceil\frac{j}{\upsilon}\rceil+1:\forall j \in \mathcal{J}_i\}, \label{eqn:K} \\
& \mathcal{N}_i=\{\Lambda_{\lceil \frac{j}{\upsilon}\rceil+1,2}(j-(\lceil\frac{j}{\upsilon}\rceil-1)\upsilon):\forall j\in \mathcal{J}_i\}. \label{eqn:N}
\end{align}Then, we have the following theorem.

\begin{theorem}\label{theorem2}
In the case of $M<N$, the minimum value of $\tau_3$ to guarantee perfect estimation of $\mv{\lambda}$ according to (\ref{eqn:received pilot phase 3}) is given by
\begin{align}\label{eqn:tau 3 case 2}
\tilde{\tau}_3=\left\lceil \frac{(K-1)N}{M} \right\rceil.
\end{align}To perfectly estimate $\mv{\lambda}$ given the above minimum value of $\tau_3$, at time slot $\tau_1+\tau_2+i$ with $1\leq i \leq (K-1)\rho$, we can set
\begin{align}
& \hspace{-6pt} a_{k,\tau_1+\tau_2+i}\hspace{-2pt}=\hspace{-2pt}\left\{\begin{array}{ll}\hspace{-2pt}1, & \hspace{-3pt} {\rm if} ~ k=\left \lceil \frac{i}{\rho}\right \rceil+1, \\ \hspace{-2pt}0, & \hspace{-3pt} {\rm otherwise},\end{array}\right. \label{eqn:pilot TDMA 1} \\
& \hspace{-6pt} \phi_{n,\tau_1+\tau_2+i}\hspace{-2pt}=\hspace{-2pt}\left\{\begin{array}{ll}\hspace{-2pt}1, & \hspace{-3pt} {\rm if} ~ n\in \Omega_i, \\ \hspace{-2pt} 0, & \hspace{-3pt} {\rm otherwise},\end{array}\right. ~~~ \hspace{-2pt} 1\hspace{-2pt}\leq \hspace{-2pt}i \hspace{-2pt}\leq \hspace{-2pt}(K-1)\rho. \label{eqn:reflecting coefficient TDMA}
\end{align}With the above solution, at each time instant $\tau_1+\tau_2+i$, we can perfectly estimate the following $\lambda_{k,n}$'s
\begin{align}
\hspace{-2pt} & [\lambda_{\lceil\frac{i}{\rho}\rceil+1,\Omega_i(1)},\cdots,\lambda_{\lceil\frac{i}{\rho}\rceil+1,\Omega_i(M)}]^T \nonumber \\ \hspace{-2pt} =&[\mv{g}_{1,\Omega_i(1)},\hspace{-2pt}\cdots\hspace{-2pt},\mv{g}_{1,\Omega_i(M)}]^{-1}\frac{\bar{\mv{y}}^{(\tau_1+\tau_2+i)}}{\sqrt{p}}, 1\hspace{-2pt}\leq\hspace{-2pt} i \hspace{-2pt}\leq \hspace{-2pt} (K\hspace{-2pt}-\hspace{-2pt}1)\rho. \label{eqn:perfect lambda case 1}
\end{align}Further, at time slot $\tau_1+\tau_2+i$ with $(K-1)\rho+1\leq i \leq \tilde{\tau}_3$, we can set
\begin{align}
&  a_{k,\tau_1+\tau_2+i}\hspace{-2pt}=\hspace{-2pt}\left\{\begin{array}{ll}\hspace{-2pt}1, & \hspace{-3pt} {\rm if} ~ k\in \mathcal{K}_i, \\ \hspace{-2pt}0, & \hspace{-3pt} {\rm otherwise},\end{array}\right. \label{eqn:pilot TDMA 2} \\
&  \phi_{n,\tau_1+\tau_2+i}\hspace{-2pt}=\hspace{-2pt}\left\{\begin{array}{ll}\hspace{-2pt}1, & \hspace{-3pt} {\rm if} ~ n\in \mathcal{N}_i, \\ \hspace{-2pt} 0, & \hspace{-3pt} {\rm otherwise},\end{array}\right. ~ \hspace{-2pt} (K-1)\rho+1\hspace{-2pt}\leq \hspace{-2pt}i \hspace{-2pt}\leq \hspace{-2pt}\tilde{\tau}_3. \label{eqn:reflecting coefficient TDMA 2}
\end{align}With the above solution, at each time instant $\tau_1+\tau_2+i$, we can perfectly estimate the following $\lambda_{k,n}$'s
\begin{align}
&[\lambda_{\lceil\frac{\mathcal{J}_i(1)}{\upsilon}\rceil+1,\mathcal{N}_i(1)},\cdots, \lambda_{\lceil\frac{\mathcal{J}_i(M_i)}{\upsilon}\rceil+1,\mathcal{N}_i(M_i)}]^T \nonumber \\ =&[\mv{g}_{1,\mathcal{N}_i(1)},\cdots,\mv{g}_{1,\mathcal{N}_i(M_i)}]^{\dag}\frac{\tilde{\mv{y}}^{(\tau_1\hspace{-1pt}+\hspace{-1pt}\tau_2\hspace{-1pt}+\hspace{-1pt}i)}}{\sqrt{p}}, (K\hspace{-2pt}-\hspace{-2pt}1)\rho\hspace{-2pt}+\hspace{-2pt}1\leq \hspace{-2pt} i \hspace{-2pt} \leq \hspace{-2pt} \tilde{\tau}_3, \label{eqn:perfect lambda case 2}
\end{align}where
\begin{align}
& M_i=|\mathcal{N}_i|, \\ & \tilde{\mv{y}}^{(\tau_1+\tau_2+i)}\hspace{-2pt}=\hspace{-2pt}\bar{\mv{y}}^{(\tau_1\hspace{-1pt}+\hspace{-1pt}\tau_2\hspace{-1pt}+\hspace{-1pt}i)}\hspace{-2pt}-\hspace{-2pt}\sum\limits_{k\in \mathcal{K}_i}\sum\limits_{n\in \mathcal{N}_i \cap \Lambda_{k,1}}\sqrt{p}\lambda_{k,n}\mv{g}_{1,n}. \label{eqn:signal after interference cancellation}
\end{align}
\end{theorem}

\begin{IEEEproof}
Please refer to Appendix \ref{appendix2}.
\end{IEEEproof}

The basic idea to achieve the minimum pilot sequence length (\ref{eqn:tau 3 case 2}) under the case of $M<N$ is as follows. First, at time slot $\tau_1+\tau_2+i$, $1\leq i \leq (K-1)\rho$, only one user, i.e., user $\lceil i /p \rceil+1$, transmits a pilot symbol 1, and $M$ IRS elements in the set of $\Omega_i$ are switched on such that the corresponding reflected channels can be perfectly estimated based on (\ref{eqn:perfect lambda case 1}). After $(K-1)\rho$ time slots, for each user $k\geq 2$, still $\upsilon<M$ $\lambda_{k,n}$'s with $n\in \Lambda_{k,2}$ are unknown. Therefore, to guarantee that at each time slot $\tau_1+\tau_2+i$, $i>(K-1)\rho$ (unless the last time slot), again $M$ $\lambda_{k,n}$'s can be estimated such that (\ref{eqn:tau 3 case 2}) is achievable, more than one user should transmit their pilots at the same time. The challenge is that if $M$ IRS elements are switched on and more than one user transmits the pilots at one time slot, the number of $\lambda_{k,n}$'s involved in the received signal is larger than $M$, as shown in (\ref{eqn:received signal 2}). Interestingly, we can show that if the user pilot and IRS reflection coefficients are designed based on (\ref{eqn:pilot TDMA 2}) and (\ref{eqn:reflecting coefficient TDMA 2}), at each time slot only $M$ $\lambda_{k,n}$'s are unknown in (\ref{eqn:received signal 2}), while the other $\lambda_{k,n}$'s have already been estimated before. As a result, after interference cancellation as shown in (\ref{eqn:signal after interference cancellation}), still $M$ $\lambda_{k,n}$'s can be perfectly estimated based on (\ref{eqn:perfect lambda case 2}) at each time slot.

To further explain the above procedure for channel estimation, we provide a simple example as follows.

\begin{example}\label{example1}
Consider the case when $M=2$, $K=3$, and $N=3$. In this case, we have $\tilde{\tau}_3=3$, and $\rho=\upsilon=1$. According to (\ref{eqn:Lambda 2}) and (\ref{eqn:Lambda 1}), we set $\Lambda_{2,1}=\{2,3\}$, $\Lambda_{2,2}=\{1\}$, $\Lambda_{3,1}=\{1,3\}$, and $\Lambda_{3,2}=\{2\}$. Then, it follows from (\ref{eqn:Omega}), (\ref{eqn:K}), and (\ref{eqn:N}) that $\Omega_1=\{2,3\}$, $\Omega_2=\{1,3\}$, $\mathcal{K}_3=\{2,3\}$, and $\mathcal{N}_3=\{1,2\}$. Based on Theorem \ref{theorem2}, the pilot sequence assigned to users $2$ and $3$ are $\mv{a}_2^{{\rm III}}=[1,0,1]^T$ and $\mv{a}_3^{{\rm III}}=[0,1,1]^T$, and the IRS reflecting coefficients are $\phi_{2,1}=\phi_{3,1}=\phi_{1,2}=\phi_{3,2}=\phi_{1,3}=\phi_{2,3}=1$ and $\phi_{n,i}=0$ otherwise. According to (\ref{eqn:perfect lambda case 1}), in time instants $\tau_1+\tau_2+1$ and $\tau_1+\tau_2+2$, we have
\begin{align}
& [\lambda_{2,2},\lambda_{2,3}]^T=[\mv{g}_{1,2},\mv{g}_{1,3}]^{-1}\bar{\mv{y}}^{\tau_1+\tau_2+1}/\sqrt{p}, \\
& [\lambda_{3,1},\lambda_{3,3}]^T=[\mv{g}_{1,1},\mv{g}_{1,3}]^{-1}\bar{\mv{y}}^{\tau_1+\tau_2+2}/\sqrt{p}.
\end{align}In time instant $\tau_1+\tau_2+3$, since $\lambda_{2,2},\lambda_{2,3},\lambda_{3,1},\lambda_{3,3}$ are already known, their interference to estimate $\lambda_{2,1}$ and $\lambda_{3,2}$ can be canceled from the received signal $\bar{\mv{y}}^{(\tau_1+\tau_2+2)}$ as shown in (\ref{eqn:signal after interference cancellation}) to get $\tilde{\mv{y}}^{(\tau_1+\tau_2+2)}$. Then, based on (\ref{eqn:signal after interference cancellation}), we have
\begin{align}
[\lambda_{2,1},\lambda_{3,2}]^T=[\mv{g}_{1,1},\mv{g}_{1,2}]^{(-1)}\tilde{\mv{y}}^{(\tau_1+\tau_2+3)}/\sqrt{p}.
\end{align}Thereby, using $\tilde{\tau}_3=3$ time instants, $\lambda_{k,n}$'s, $k=2,3$, $n=1,2,3$, are all perfectly estimated based on Theorem \ref{theorem2}.
\end{example}

According to Theorems \ref{theorem1} and \ref{theorem2}, we manage to reduce the channel estimation time duration in Phase III from $(K-1)N$ symbols to
\begin{align}\label{eqn:tau 3 case 2 all}
\tilde{\tau}_3=\max\left(K-1,\left\lceil \frac{(K-1)N}{M} \right\rceil\right),
\end{align}symbols thanks to the hidden relation shown in (\ref{eqn:correlated channel}). Further, the designs of user pilot and IRS reflecting coefficients shown in Theorems \ref{theorem1} and \ref{theorem2} are independent of $\mv{g}_{1,n}$'s. Thereby, channel estimation in Phase III does not require any channel feedback from the BS to the users and IRS.

\subsection{Overall Channel Estimation Overhead}

To summarize, for perfectly estimating all the direct channels $\mv{h}_k$'s and reflected channels $\mv{g}_{k,n}$'s in the case without noise at the BS, the minimum pilot sequence length is
\begin{align}\label{eqn:total channel estimation time}
\tilde{\tau}\hspace{-2pt}=\hspace{-2pt}\tilde{\tau}_1\hspace{-2pt}+\hspace{-2pt}\tilde{\tau}_2\hspace{-2pt}+\hspace{-2pt}\tilde{\tau}_3\hspace{-2pt}=\hspace{-2pt}K\hspace{-2pt}+\hspace{-2pt}N\hspace{-2pt}+\hspace{-2pt}\max\left(K\hspace{-2pt}-\hspace{-2pt}1,\left \lceil\frac{(K\hspace{-2pt}-\hspace{-2pt}1)N}{M} \right \rceil \right).
\end{align}

Interestingly, in the massive MIMO regime \cite{Larsson14}, i.e., $M\rightarrow \infty$, $\tilde{\tau}$ reduces to
\begin{align}\label{eqn:total channel estimation time massive MIMO}
\tilde{\tau}=K+N+K-1=2K+N-1,
\end{align}which is linear with $K$ and $N$. Thereby, under the three-phase channel estimation protocol for IRS-assisted uplink communications, massive MIMO makes it possible to effectively estimate $KMN+KM$ unknown channel coefficients using a scalable number of pilot symbols. Such a result is in sharp contrast to the traditional channel estimation scenario without IRS, where the minimum channel estimation time does not depend on the number of receive antennas \cite{Hassibi03}.

\section{Channel Estimation for Case with Noise}\label{sec:Channel Estimation in Case with Noise}
In the previous section, we have shown how to perfectly estimate all the channels using at least $\tilde{\tau}$ time instants for the ideal case without noise at the BS, where $\tilde{\tau}$ is given in (\ref{eqn:total channel estimation time}). In this section, we introduce how to estimate the channels under our proposed three-phase channel estimation protocol for the case with noise at the BS, using $\tau_1\geq \tilde{\tau}_1$, $\tau_2\geq \tilde{\tau}_2$, and $\tau_3\geq \tilde{\tau}_3$ time slots in Phases I, II, and III, respectively.

\subsection{Phase I: Direct Channel Estimation}\label{sec3}
With noise at the BS, the received signal in Phase I given in (\ref{eqn:received pilot direct channel}) is re-expressed as
\begin{align}\label{eqn:received pilot direct channel noise}
\mv{Y}^{{\rm I}} =\sqrt{p}[\mv{h}_1,\cdots,\mv{h}_K][\mv{a}_1^{{\rm I}},\ldots,\mv{a}_K^{{\rm I}}]^T+\mv{Z}^{{\rm I}},
\end{align}where $\mv{Z}^{{\rm I}}=[\mv{z}^{(1)},\cdots,\mv{z}^{(\tau_1)}]$. According to \cite{Hassibi03}, the optimal pilot design should guarantee that the pilot sequences of different users are orthogonal with each other, i.e., $(\mv{a}_k^{{\rm I}})^T\mv{a}_j^{{\rm I}}=0$, $\forall k\neq j$. With $\tau_1\geq K$, we can set the pilot sequence of user $k$ as
\begin{align}\label{eqn:pilot phase 1 noise}
\mv{a}_k^{{\rm I}}=[1,e^{\frac{-2(k-1)\pi q}{\tau_1}},\cdots,e^{\frac{-2(k-1)(\tau_1-1)\pi q}{\tau_1}}]^T, ~~~ \forall k,
\end{align}where $q^2=-1$. In this case, the minimum mean-squared error (MMSE) channel estimator is
\begin{align}\label{eqn:channel estimation phase 1 noise}
\hat{\mv{h}}_k=\frac{\beta_k^{{\rm BU}}\sqrt{p}}{\beta_k^{{\rm BU}}p\tau_1+\sigma^2}\mv{Y}^{{\rm I}}\mv{a}_k^\ast, ~~~ \forall k.
\end{align}Further, the MSE for estimating $\mv{h}_k$ is denoted by
\begin{align}\label{eqn:MSE phase 1}
\varepsilon_k^{{\rm I}}\hspace{-2pt}=\hspace{-2pt}\mathbb{E}_{\mv{h}_k}[(\hat{\mv{h}}_k\hspace{-2pt}-\hspace{-2pt}\mv{h}_k)^H(\hat{\mv{h}}_k\hspace{-2pt}-\hspace{-2pt}\mv{h}_k)]=\frac{M\beta_k^{{\rm BU}}\sigma^2}{\beta_k^{{\rm BU}}p\tau_1\hspace{-2pt}+\hspace{-2pt}\sigma^2}, ~ \forall k.
\end{align}

\subsection{Phase II: Reflecting Channel Estimation for Typical User}\label{sec4}
In Phase II, the typical user 1 transmits the pilot symbols to the BS, which will cancel the interference caused by $\mv{h}_k$'s for estimating $\mv{g}_{1,n}$'s. In the case with noise the BS, the effective received signal at the BS after interference cancellation given in (\ref{eqn:received pilot reflected channel user 1}) is re-expressed as\begin{align}
\bar{\mv{y}}^{(i)}&=\mv{y}^{(i)}-\hat{\mv{h}}_1\sqrt{p}a_{1,i} \nonumber \\ & = \sum_{n=1}^{N}\phi_{n,i}\mv{g}_{1,n}\sqrt{p}a_{1,i}+(\mv{h}_1-\hat{\mv{h}}_1)\sqrt{p}a_{1,i}+\mv{z}^{(i)}.
\end{align}Note that due to the imperfect channel estimation in Phase I, $\mv{h}_1-\hat{\mv{h}}_1\neq \mv{0}$ in general. Then, the overall effective received signals over Phase II is given by
\begin{align}\label{eqn:received pilot phase 2 noise}
\bar{\mv{Y}}^{{\rm II}}
= & \sqrt{p}[\mv{g}_{1,1},\cdots,\mv{g}_{1,N}]\mv{\Phi}^{{\rm II}}{\rm diag}(\mv{a}_1^{{\rm II}})\nonumber \\ & +\sqrt{p}(\mv{h}_1-\hat{\mv{h}}_1)(\mv{a}_1^{{\rm II}})^T+\mv{Z}^{{\rm II}},
\end{align}where $\mv{Z}^{{\rm II}}=[\mv{z}^{(\tau_1+1)},\cdots,\mv{z}^{(\tau_1+\tau_2)}]$. For convenience, define $\mv{C}^{{\rm BI}}_1=\mathbb{E}[[\mv{g}_{1,1},\cdots,\mv{g}_{1,N}]^H[\mv{g}_{1,1},\cdots,\mv{g}_{1,N}]]$, and $\bar{\mv{Z}}^{{\rm II}}=\sqrt{p}(\mv{h}_1-\hat{\mv{h}}_1)(\mv{a}_1^{{\rm II}})^T+\mv{Z}^{{\rm II}}$ as the overall noise consisting of channel estimation error in Phase I and AWGN. Then, we have
\begin{align}
\hspace{-5pt}\mv{\Psi}_{{\rm II}}\hspace{-3pt}=\hspace{-3pt}\mathbb{E}[(\bar{\mv{Z}}^{{\rm II}})^H\bar{\mv{Z}}^{{\rm II}}]\hspace{-3pt}=\hspace{-3pt}\frac{pM\beta_1^{{\rm BU}}\sigma^2}{\beta_1^{{\rm BU}}p\tau_1\hspace{-2pt}+\hspace{-2pt}\sigma^2}(\mv{a}_1^{{\rm II}})^\ast(\mv{a}_1^{{\rm II}})^T\hspace{-3pt}+\hspace{-3pt}M\sigma^2\mv{I}.
\end{align}

Since $\mv{g}_{1,n}$'s do not follow the Rayleigh fading channel model, it is difficult to design the MMSE estimator based on (\ref{eqn:received pilot phase 2 noise}). In this paper, we consider the LMMSE estimator in Phase II. By setting the pilot sequence of user 1 as
\begin{align}\label{eqn:pilot phase 2 noise}
\mv{a}_1^{{\rm II}}=[1,\cdots,1]^T,
\end{align}and the IRS reflection coefficients $\mv{\Phi}^{{\rm II}}$ as (\ref{eqn:DFT}) similar to \cite{Zheng19,Jensen}, the LMMSE channel estimator in Phase II is
\begin{align}\label{eqn:estimated channel Phase II noise}
\hspace{-5pt} & [\hat{\mv{g}}_{1,1},\cdots,\hat{\mv{g}}_{1,N}] \nonumber \\ \hspace{-5pt} =\hspace{-1pt} & \hspace{-1pt}\sqrt{p}\bar{\mv{Y}}^{{\rm II}}\mv{\Psi}_{{\rm II}}^{-1}(\mv{\Phi}^{{\rm II}})^H\left(p\mv{\Phi}^{{\rm II}}\mv{\Psi}_{{\rm II}}^{-1}(\mv{\Phi}^{{\rm II}})^H\hspace{-2pt}+\hspace{-2pt}(\mv{C}^{{\rm BI}}_1)^{-1}\right)^{-1}.
\end{align}The MSE of the above LMMSE channel estimator is
\begin{align}\label{eqn:MSE}
\varepsilon^{{\rm II}}={\rm tr}\left(\left(p\mv{\Phi}^{{\rm II}}\mv{\Psi}_{{\rm II}}^{-1}(\mv{\Phi}^{{\rm II}})^H+(\mv{C}^{{\rm BI}}_1)^{-1}\right)^{-1}\right).
\end{align}

\subsection{Phase III: Reflecting Channel Estimation for Other Users}
In Phase III, with the imperfect estimation of $\mv{h}_k$'s and $\mv{g}_{1,n}$'s, the effective received signal at time slot $i$, $i=\tau_1+\tau_2+1,\cdots,\tau_1+\tau_2+\tau_3$, given in (\ref{eqn:received pilot reflected channel user k 2}) can be expressed as
\begin{align}\label{eqn:imperfect received signal phase 3}
\bar{\mv{y}}^{(i)}=&\sum\limits_{k=2}^K\sum_{n=1}^{N}\phi_{n,i}\hat{\mv{g}}_{1,n}\sqrt{p}a_{k,i}\lambda_{k,n}\hspace{-2pt}+\hspace{-2pt}\sum\limits_{k=2}^K(\mv{h}_k\hspace{-2pt}-\hspace{-2pt}\hat{\mv{h}}_k)\sqrt{p}a_{k,i}\hspace{-2pt}+\hspace{-2pt}\mv{z}^{(i)} \nonumber \\ &+\sum\limits_{k=2}^K\sum_{n=1}^{N}\phi_{n,i}(\mv{g}_{1,n}-\hat{\mv{g}}_{1,n})\sqrt{p}a_{k,i}\lambda_{k,n}.
\end{align}However, it is difficult to design the LMMSE channel estimator based on (\ref{eqn:imperfect received signal phase 3}) due to the fact that $\lambda_{k,n}$'s also contribute to the noise for estimating themselves with imperfect estimation of $\mv{g}_{1,n}$'s.

To tackle the above challenge, in the following we assume that the channel estimation error $\mv{g}_{1,n}-\hat{\mv{g}}_{1,n}=\mv{0}$, $\forall n$, such that there is no self-interference generated by $\lambda_{k,n}$'s. In practice, we can increase the pilot sequence length in Phase II, i.e., $\tau_2$, such that the estimation of $\mv{g}_{1,n}$'s is sufficiently accurate. Under the above assumption, the effective received signal given in (\ref{eqn:imperfect received signal phase 3}) reduces to
\begin{align}\label{eqn:imperfect received signal III}
\tilde{\mv{y}}^{(i)}\hspace{-3pt}=\hspace{-3pt}\sum\limits_{k=2}^K\sum_{n=1}^{N}\phi_{n,i}\mv{g}_{1,n}\sqrt{p}a_{k,i}\lambda_{k,n}\hspace{-2pt}+\hspace{-2pt}\sum\limits_{k=2}^K(\mv{h}_k\hspace{-2pt}-\hspace{-2pt}\hat{\mv{h}}_k)\sqrt{p}a_{k,i}\hspace{-2pt}+\hspace{-2pt}\mv{z}^{(i)}.
\end{align}

Moreover, we assume an orthogonal transmission and reflection strategy: at each time slot $i$, only one user, denoted by user $k_i$, transmits its pilot symbol to the BS, and only $M_i\leq M$ out of $N$ IRS elements, denoted by the set $\Delta_i$ with $M_i=|\Delta_i|\leq M$, are switched on to reflect the pilot symbol, such that the BS can estimate $\mv{g}_{k_i,n}$'s, $\forall n\in \Delta_i$. Specifically, with $i\geq \tau_1+\tau_2+1$, we define
\begin{align}
& k_i=\left\lceil \frac{i-\tau_1-\tau_2}{\lceil N/M\rceil}\right\rceil+1, \label{eqn:TDMA user} \\
& \Delta_i=\left\{\begin{array}{ll} \{\varphi_i+1,\cdots,\varphi_i+M\},\hspace{-1pt} & \hspace{-1pt}{\rm if} ~ \left\lfloor \frac{i\hspace{-1pt}-\hspace{-1pt}\tau_1\hspace{-1pt}-\hspace{-1pt}\tau_2}{\lceil N/M \rceil}\right\rfloor\hspace{-2pt}\neq\hspace{-2pt} \frac{i\hspace{-1pt}-\hspace{-1pt}\tau_1\hspace{-1pt}-\hspace{-1pt}\tau_2}{\lceil N/M \rceil}, \\ \{(\lceil \frac{N}{M} \rceil\hspace{-2pt}-\hspace{-2pt}1)M\hspace{-2pt}+\hspace{-2pt}1,\cdots,N\}, \hspace{-1pt}&\hspace{-1pt} {\rm otherwise}. \end{array} \right. \label{eqn:TDMA IRS}
\end{align}where
\begin{align}
\varphi_i=\left(i-\tau_1-\tau_2-\left\lfloor \frac{i-\tau_1-\tau_2}{\lceil N/M \rceil}\right\rfloor\lceil N/M \rceil-1\right)M.
\end{align}As a result, for user $k_i$, $\lceil N/M \rceil$ time slots are allocated to estimate its reflected channels. In each of the first $\lceil N/M \rceil-1$ time slots, $M$ IRS elements are switched on, while in the last time slot, the remaining $N-(\lceil N/M \rceil-1)M$ IRS elements are switched on. In total, under this orthogonal transmission and reflection strategy, $\tau_3=(K-1)\lceil N/M \rceil$ time slots are needed to estimate $\mv{g}_{k,n}$'s, $\forall k\geq 2$, $\forall n$.

Next, we show how to estimate $\mv{g}_{k,n}$'s based on the above orthogonal transmission and reflection strategy. At time slot $i$, $i=\tau_1+\tau_2+1,\cdots,\tau_1+\tau_2+(K-1)\lceil N/M \rceil$, we set
\begin{align}
& a_{k,i}=\left\{\begin{array}{ll}1, & {\rm if} ~ k= k_i, \\ 0, & {\rm otherwise}, \end{array}\right. ~ \forall k\geq 2,\label{eqn:pilot phase 3 noise} \\
& \phi_{n,i}=\left\{\begin{array}{ll}1, & {\rm if} ~ n\in \Delta_i, \\ 0, & {\rm otherwise}, \end{array}\right. ~ \forall n. \label{eqn:reflection coefficient phase 3 noise}
\end{align}Then, the effective received signal at time slot $i$ given in (\ref{eqn:imperfect received signal III}) reduces to
\begin{align}
\tilde{\mv{y}}^{(i)}=&\sum_{n\in \Delta_i}\mv{g}_{1,n}\sqrt{p}\lambda_{k_i,n}+(\mv{h}_{k_i}-\hat{\mv{h}}_{k_i})\sqrt{p}+\mv{z}^{(i)} \nonumber \\
= &\sqrt{p} \mv{G}_{1,i}\mv{\lambda}_{k_i,i}+\tilde{\mv{z}}^{(i)},
\end{align}where
\begin{align}
& \mv{G}_{1,i}=[\mv{g}_{1,\Delta_i(1)},\cdots,\mv{g}_{1,\Delta_i(|\Delta_i|)}], \\
& \mv{\lambda}_{k_i,i}=[\lambda_{k_i,\Delta_i(1)},\cdots,\lambda_{k_i,\Delta_i(|\Delta_i|)}]^T, \\
& \tilde{\mv{z}}^{(i)}=\sqrt{p}(\mv{h}_{k_i}-\hat{\mv{h}}_{k_i})+\mv{z}^{(i)},
\end{align}with $\Delta_i(j)$ denoting the $j$th element of $\Delta_i$. Next, define $\mv{C}^{\lambda}_i=\mathbb{E}[\mv{\lambda}_{k_i,i}\mv{\lambda}_{k_i,i}^H]$, and
\begin{align}
\mv{\Psi}_{{\rm III}}=&\mathbb{E}[\tilde{\mv{z}}^{(i)}(\tilde{\mv{z}}^{(i)})^H] \nonumber \\=&\frac{\beta_{k_i}^{{\rm BU}}p \sigma^4}{\beta_{k_i}^{{\rm BU}}p\tau_1+\sigma^2}\mv{C}^{{\rm B}}_{k_i}+\frac{(\beta_{k_i}^{{\rm BU}}p)^2\tau_1 \sigma^2}{\beta_{k_i}^{{\rm BU}}p\tau_1+\sigma^2}\mv{I}+\sigma^2\mv{I}.
\end{align}Given any realization of $\mv{G}_{1,i}$ that is assumed to be perfectly estimated in Phase II, the LMMSE channel estimator in Phase III is then given by
\begin{align}\label{eqn:LMMSE Phase 3}
\hat{\mv{\lambda}}_{k_i,i}(\mv{G}_{1,i})\hspace{-2pt}=\hspace{-2pt}\sqrt{p}\left(p\mv{G}_{1,i}^H\mv{\Psi}_{{\rm III}}^{-1}\mv{G}_{1,i}\hspace{-2pt}+\hspace{-2pt}(\mv{C}_i^{\lambda})^{-1}\right)^{-1}\hspace{-2pt}\mv{G}_{1,i}^H\mv{\Psi}_{{\rm III}}^{-1}\tilde{\mv{y}}^{(i)}.
\end{align}The corresponding MSE for estimating $\lambda_{k_i,i}$ given $\mv{G}_{1,i}$ is
\begin{align}
\varepsilon_{k_i,i}^{{\rm III}}(\mv{G}_{1,i})=&\mathbb{E}[\|\hat{\mv{\lambda}}_{k_i,i}(\mv{G}_{1,i})-\mv{\lambda}_{k_i,i}\|^2|\mv{G}_{1,i}]\nonumber \\ =&{\rm tr}\left( \left(p\mv{G}_{1,i}^H\mv{\Psi}_{{\rm III}}^{-1}\mv{G}_{1,i}+(\mv{C}_i^{\lambda})^{-1}\right)^{-1}\right).
\end{align}By averaging all the channel realizations of $\mv{G}_{1,i}$'s, the MSE for estimating $\lambda_{k_i,i}$ is
\begin{align}
\varepsilon_{k_i,i}^{{\rm III}}\hspace{-2pt}=\hspace{-2pt}\mathbb{E}_{\mv{G}_{1,i}}\left[{\rm tr}\left(\left(p\mv{G}_{1,i}^H\mv{\Psi}_{{\rm III}}^{-1}\mv{G}_{1,i}\hspace{-2pt}+\hspace{-2pt}(\mv{C}_i^{\lambda})^{-1}\right)^{-1}\right)\right].
\end{align}

At last, the overall MSE for estimating $\lambda_{k_i,i}$'s $\forall i$ is given by $\varepsilon^{{\rm III}}=\sum_{i=1}^{\tau_3}\varepsilon_{k_i,i}^{{\rm III}}$.

\subsection{Overall Channel Estimation Strategy}
For the case with noise at the BS, the overall channel estimation strategy is summarized in Table \ref{table3}.

\begin{table}[htp]
\begin{center}
\caption{\textbf{Algorithm \ref{table3}}: Overall Channel Estimation Strategy} \vspace{0.2cm}
 \hrule
\vspace{0.2cm}
\begin{itemize}
\item[1.] {\bf Phase I} ($i=1,\cdots,\tau_1$): The users transmit pilot symbols $\mv{a}_k^{{\rm I}}$'s based on (\ref{eqn:pilot phase 1 noise}). The IRS is switched off, i.e., $\phi_{n,i}$'s are set to be zero. The BS estimates the direct channels $\mv{h}_k$'s according to (\ref{eqn:channel estimation phase 1 noise});
\item[2.] {\bf Phase II} ($i=\tau_1+1,\cdots,\tau_1+\tau_2$): Only user 1 transmits non-zero pilot symbols $\mv{a}_1^{{\rm II}}$ based on (\ref{eqn:pilot phase 2 noise}). The IRS is switched on, and $\phi_{n,i}$'s are set based on (\ref{eqn:DFT}). The BS removes the interference caused by $\mv{h}_k$'s based on (\ref{eqn:received pilot phase 2 noise}) and estimates $\mv{g}_{1,n}$'s based on (\ref{eqn:estimated channel Phase II noise});
\item[3.] {\bf Phase III ($i=\tau_1+\tau_2+1,\cdots,\tau_1+\tau_2+\tau_3$)}: User 2 to user $K$ transmit pilot symbols $\mv{a}_k^{{\rm III}}$'s,, $k\geq 2$, based on (\ref{eqn:pilot phase 3 noise}). The IRS is switched on, and $\phi_{n,i}$'s are set as (\ref{eqn:reflection coefficient phase 3 noise}). The BS removes the interference caused by $\mv{h}_k$'s based on (\ref{eqn:imperfect received signal III}) and estimates $\mv{\lambda}$ based on (\ref{eqn:LMMSE Phase 3}).
\end{itemize}
\vspace{0.2cm} \hrule \label{table3}\vspace{-10pt}
\end{center}
\end{table}

\section{Numerical Examples}\label{sec:Numerical Examples}
In this section, we provide numerical examples to verify the effectiveness of our proposed three-phase channel estimation protocol in the IRS-assisted multiuser communications. We assume that the IRS is equipped with $N=32$ reflecting elements. Moreover, the path loss of $\mv{h}_k$'s, $t_{k,n}$'s, and $\mv{r}_n$'s is modeled as $\beta_k^{{\rm BU}}=\beta_0(d_k^{{\rm BU}}/d_0)^{-\alpha_1}$, $\beta_k^{{\rm IU}}=\beta_0(d_k^{{\rm IU}}/d_0)^{-\alpha_2}$, and $\beta^{{\rm BI}}=\beta_0(d^{{\rm BI}}/d_0)^{-\alpha_3}$, respectively, where $d_0=1$ meter (m) denotes the reference distance, $\beta_0=-20$ dB denotes the path loss at the reference distance, $d_k^{{\rm BU}}$, $d_k^{{\rm IU}}$, and $d^{{\rm BI}}$ denote the distance between the BS and user $k$, between the IRS and user $k$, as well as between the BS and the IRS, while $\alpha_1$, $\alpha_2$, and $\alpha_3$ denote the path loss factors for $\mv{h}_k$'s, $t_{k,n}$'s, and $\mv{r}_n$'s, respectively. In the numerical examples, we set $\alpha_1=4.2$, $\alpha_2=2.1$, and $\alpha_3=2.2$. The distance between the BS and IRS is set to be $d^{{\rm BI}}=100$ m, and all the users are assumed to be located in a circular regime, whose center is $10$ m away from the IRS and $105$ m away from the BS, and radius is $5$ m. For the correlation matrix $\mv{C}_k^{{\rm B}}$, we adopt the exponential correlation matrix model \cite{Loyka01}, where the element in the $i$th row and $j$th column is denoted by $[\mv{C}_k^{{\rm B}}]_{i,j}=(c_k^{{\rm B}})^{i-j}$ if $i\geq j$, and $[\mv{C}_k^{{\rm B}}]_{i,j}=[\mv{C}_k^{{\rm B}}]_{j,i}^\ast$, if $i<j$, with $|c_k^{{\rm B}}|<1$. Similarly, the correlation matrices $\mv{C}^{{\rm B}}$, $\mv{C}^{{\rm I}}$, and $\mv{C}_k^{{\rm I}}$ follow the exponential correlation matrix model as well. The identical transmit power of users is $33$ dBm. The channel bandwidth is assumed to be $1$ MHz, and the power spectrum density of the AWGN at the BS is $-169$ dBm/Hz. Moreover, to illustrate the performance gain of our proposed channel estimation framework, we consider the following benchmark scheme.

{\bf Benchmark Scheme:} In Phases I and II, the estimation of $\mv{h}_k$'s and $\mv{g}_{1,n}$'s is the same as the proposed framework. However, in Phase III, from time slots $\tau_1+\tau_2+(k-2)\tau_2+1$ to $\tau_1+\tau_2+(k-1)\tau_2$, $\forall k\geq 2$, only user $k$ transmits its $\tau_2$ pilot symbols to the BS, while the estimation of $\mv{g}_{k,n}$, $\forall n$, is the same as that of $\mv{g}_{1,n}$, $\forall n$, for both cases without and with noise at the BS. In other words, the channel correlations in (\ref{eqn:correlated channel}) is not exploited to improve the channel estimation performance.

In the following, we provide numerical examples to evaluate the performance of the proposed channel estimation framework for the cases without and with noise at the BS, respectively.

\subsection{The Case without Noise at the BS}
\begin{figure}[t]
  \centering
  \includegraphics[width=9cm]{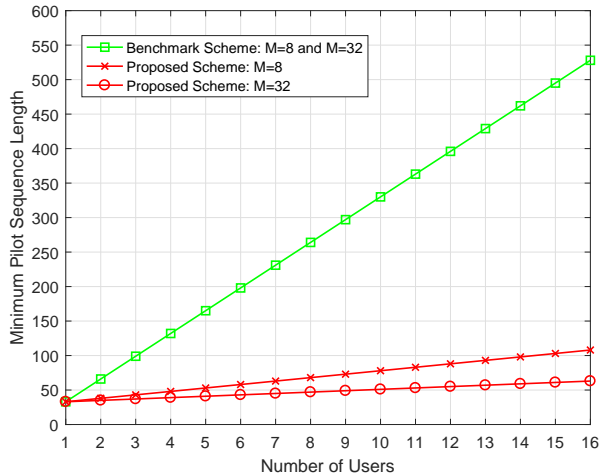}\vspace{-5pt}
  \caption{Minimum pilot sequence length versus number of users.}\label{Fig3}\vspace{-10pt}
\end{figure}

First, we illustrate the minimum estimation time performance of the proposed framework in the case without noise at the BS. Fig. \ref{Fig3} shows the minimum pilot sequence length for perfect channel estimation as given in (\ref{eqn:total channel estimation time}) versus the number of users when the number of antennas at the BS is $M=8$ and $M=32$, respectively. Moreover, the minimum pilot sequence length required by the benchmark scheme, which is characterized as $K+KN$, is also shown. It is observed that thanks to the exploitation of the channel correlation (\ref{eqn:correlated channel}), the minimum pilot sequence length under our proposed framework increases much more slowly with the number of users than that under the benchmark scheme. Moreover, by comparing the cases when $M=8$ and $M=32$, it is observed that under our proposed framework, the minimum pilot sequence length decreases very fast as the number of BS antennas increases; while that under the benchmark strategy is independent of the number of BS antennas.

\subsection{The Case with Noise at the BS}
\begin{figure}[t]
  \centering
  \includegraphics[width=9cm]{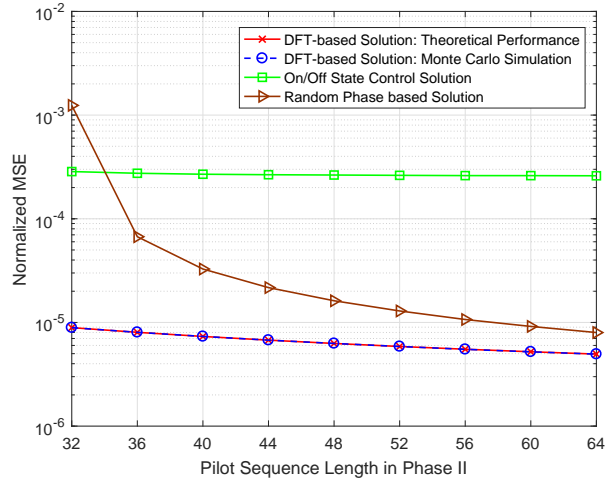}\vspace{-5pt}
  \caption{Normalized MSE comparison in Phase II.}\label{Fig5}\vspace{-10pt}
\end{figure}

Next, we show the MSE performance of the proposed framework in the case with noise at the BS. In the following, we assume that the number of users is $8$, and the BS is equipped with $32$ antennas. First, we consider the performance of Phase II for estimating $\mv{g}_{1,n}$'s, $\forall n$. Besides the proposed DFT-based solution shown in (\ref{eqn:DFT}), we also consider two more schemes for performance comparison. Specifically, in the first scheme, only one IRS element is switched on at each time slot, and all the IRS elements take turns to be in the ``on'' state. In the second scheme, the phase shifter of each IRS element at each time slot is randomly chosen in the range of $[0,2\pi)$. Fig. \ref{Fig5} shows the normalized MSE in Phase II achieved by different schemes, which is defined as
\begin{align}
e^{{\rm II}}=\frac{\sum\limits_{n=1}^N\mathbb{E}[\|\hat{\mv{g}}_{1,n}-\mv{g}_{1,n}\|^2]}{\sum\limits_{n=1}^N\mathbb{E}[\|\mv{g}_{1,n}\|^2]}.
\end{align}
First, it is observed that our theoretical characterization of channel estimation MSE, i.e., (\ref{eqn:MSE}), matches the Monte Carlo simulation perfectly. Second, it is observed that the normalized MSE under the DFT-based solution is much smaller than that under the other two schemes.

\begin{figure}[t]
  \centering
  \includegraphics[width=9cm]{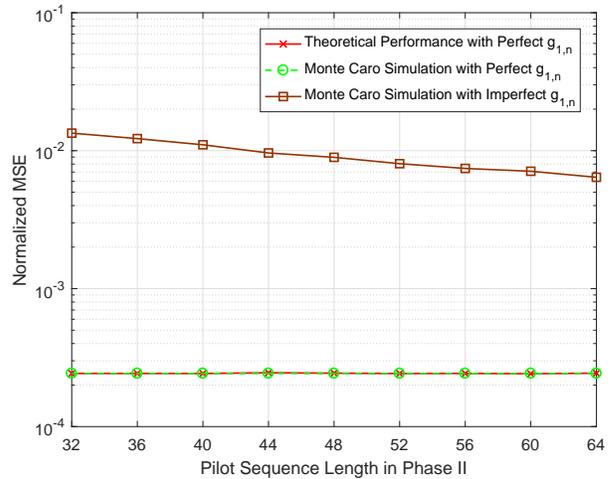}\vspace{-5pt}
  \caption{Normalized MSE in Phase III: theoretical results versus Monte Carlo simulations.}\label{Fig6}\vspace{-10pt}
\end{figure}

Next, we evaluate the performance for estimating $\lambda_{k,n}$'s in Phase III. Similar to Phase II, we define the normalized MSE for estimating $\lambda_{k,n}$'s in Phase III as
\begin{align}
e^{{\rm III}}=\frac{\sum\limits_{k=2}^K\sum\limits_{n=1}^N\mathbb{E}[|\hat{\lambda}_{k,n}-\lambda_{k,n}|^2]}{\sum\limits_{k=2}^K\sum\limits_{n=1}^N\mathbb{E}[|\lambda_{k,n}|^2]}.
\end{align}Fig. \ref{Fig6} shows the normalized MSE for estimating $\lambda_{k,n}$'s versus the channel training time in Phase II. It is observed that if the channel estimation of $\mv{g}_{1,n}$'s is perfect, then our theoretical characterization of the MSE for estimating $\lambda_{k,n}$'s matches the Monte Carlo simulation perfectly. In practice, the estimation of $\mv{g}_{1,n}$'s is imperfect. In this case, a small level of mismatch exists between our theoretical result under the assumption of perfect estimation of $\mv{g}_{1,n}$'s and the Monte Carlo simulation. Nevertheless, this gap can be reduced by increasing the channel estimation time in Phase II such that the estimation of $\mv{g}_{1,n}$'s is more accurate.

\begin{figure}[t]
  \centering
  \includegraphics[width=9cm]{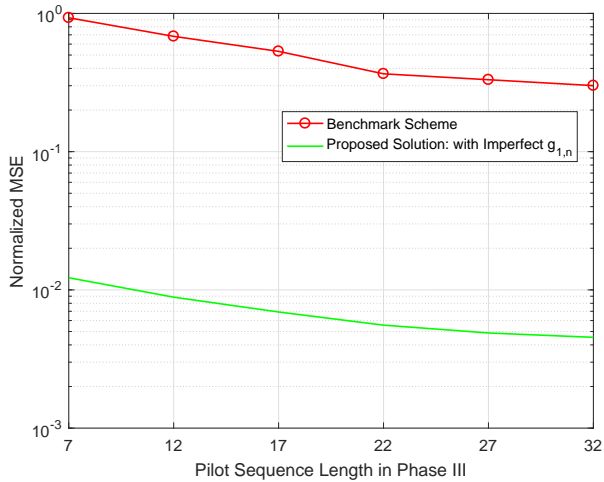}\vspace{-5pt}
  \caption{Normalized MSE comparison in Phase III.}\label{Fig7}\vspace{-10pt}
\end{figure}

Fig. \ref{Fig7} shows the MSE performance in Phase III achieved by the proposed scheme and the benchmark scheme described at the beginning of this section versus the channel estimation time in Phase III. It is observed that under the benchmark scheme, the normalized MSE is above 0.3 when $\tau_3$ ranges from $7$ to $32$. This indicates that the channel estimation error is almost as strong as the channel power. This is because the minimum pilot sequence length for perfect channel estimation in the case without noise at the base is $224$ under the benchmark strategy. As a result, when $\tau_3$ ranges from $7$ to $32$, the benchmark scheme cannot estimate the channels very well even without noise at the BS. On the other hand, under our proposed framework, the normalized MSE for estimating all the channels is below $10^{-2}$ when $\tau_3\geq 12$.

At last, we consider the overall MSE performance for estimating both $\mv{h}_k$'s and $\mv{g}_{k,n}$'s. In this case, the normalized MSE is defined as
\begin{align}
e=\frac{\sum\limits_{k=1}^K\mathbb{E}[\|\hat{\mv{h}}_k-\mv{h}_k\|^2]+\sum\limits_{k=1}^K\sum\limits_{n=1}^N\mathbb{E}[\|\hat{\mv{g}}_{k,n}-\mv{g}_{k,n}\|^2]}{\sum\limits_{k=1}^K\mathbb{E}[\|\mv{h}_k\|^2]+\sum\limits_{k=1}^K\sum\limits_{n=1}^N\mathbb{E}[\|\mv{g}_{k,n}\|^2]}.
\end{align}

\begin{figure}[t]
  \centering
  \includegraphics[width=9cm]{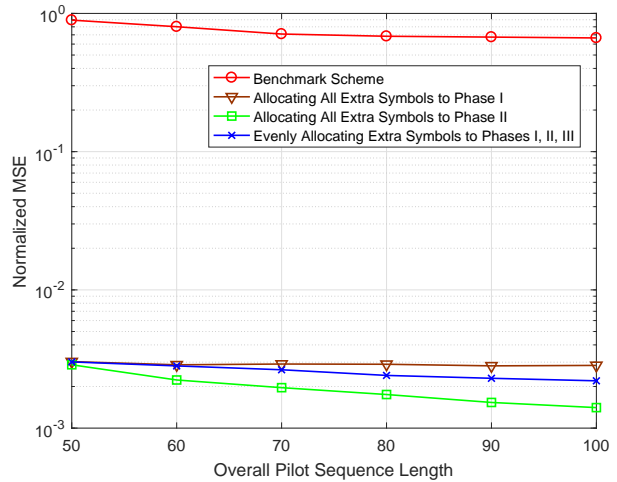}\vspace{-5pt}
  \caption{Overall normalized MSE comparison between the proposed framework and the benchmark scheme.}\label{Fig8}\vspace{-10pt}
\end{figure}

Fig. \ref{Fig8} shows the normalized MSE for estimating $\mv{h}_k$'s and $\mv{g}_{k,n}$'s achieved by various strategies. It is observed that similar to Fig. \ref{Fig7}, the proposed scheme significantly improves the overall channel estimation MSE performance compared to the benchmark scheme. Moreover, under the proposed framework, we also consider three schemes to allocate the extra time slots among the three phases if the overall channel estimation time is larger than (\ref{eqn:total channel estimation time}). Under these three schemes, the extra time slots are all allocated to Phase I, all allocated to Phase II, and evenly allocated to Phases I, II, and III, respectively. It is observed in Fig. \ref{Fig8} that the best way is to allocate more time slots to Phase II such that the error propagation to Phase III is reduced.

\section{Conclusion and Future Work}\label{sec:Conclusion}
In this paper, we proposed an innovative three-phase framework to estimate a large number of channel coefficients in the IRS-assisted uplink multiuser communications accurately using merely a small number of pilot symbols. Such an interesting result is enabled by exploiting the correlations among the IRS reflected channels: each IRS reflecting element reflects the signals from different users to the BS via the same channel. Under the proposed framework, the minimum pilot sequence length for perfect channel estimation in the case without noise at the BS was characterized, and the LMMSE channel estimators in the case with noise at the BS were derived. Numerical examples were provided to verify the effectiveness of our framework compared to the benchmark scheme without taking advantage of channel correlation.

There are a number of directions along which the channel estimation framework proposed in this paper can be further enriched. First, in Phase II of our proposed protocol, $N$ time slots are required for channel estimation of the typical user, which can be long in practice if the IRS is equipped with a large number of IRS elements. It is thus interesting to study how to make use of the channel property in IRS-assisted communication systems to reduce the channel estimation time in Phase II. Second, the channel estimation error in Phases I and II under the proposed framework will affect channel estimation in Phase III. Intuitively, if more time slots are allocated to Phase I and Phase II, the error propagation to Phase III will be reduced, while less time is left for channel estimation. As a result, how to allocate the available time slots among the three channel estimation phases is also an open problem. Similarly, it is crucial to study the pilot power allocation strategy among the three phases of our proposed framework to improve the channel estimation performance. Moreover, it is interesting to extend our channel estimation protocol to a multi-cell communication system and study the effect of pilot contamination on the system performance. Another important direction is to study whether the channel reciprocity holds in the IRS-assisted communications. If not, new approach should be proposed to utilize the channel redundancy for reducing the channel estimation time in the downlink. Last but not least, it is crucial to characterize the user achievable rate with imperfect CSI under the proposed framework.

\begin{appendix}
\subsection{Proof of Theorem \ref{theorem1}}\label{appendix1}
In the case of $M\geq N$, we first prove that there exists a unique solution to (\ref{eqn:received pilot phase 3}) only if $\tau_3 \geq K-1$. Define
\begin{align}\label{eqn:eta}
\eta_{n,i}=\sum\limits_{k=2}^K\lambda_{k,n}\phi_{n,i+\tau_1+\tau_2}a_{k,i+\tau_1+\tau_2}, ~ 1\leq n \leq N, 1\leq i \leq \tau_3.
\end{align}Then, it can be shown that (\ref{eqn:received pilot phase 3}) can be expressed as
\begin{align}\label{eqn:solution 1}
\sum\limits_{n=1}^N\eta_{n,i}\mv{g}_{1,n}= \bar{\mv{y}}^{(\tau_1+\tau_2+i)}, ~~~ i=1,\cdots,\tau_3.
\end{align}Under the channel model of $\mv{r}_n$'s and $t_{k,n}$'s shown in (\ref{eqn:IRS to BS channel}) and (\ref{eqn:user to IRS channel}), in the case of $M\geq N$, $\mv{g}_{1,n}$'s are linearly independent with each other with probability one. As a result, $\eta_{n,i}$'s, $n=1,\cdots,N$, $i=1,\cdots,\tau_3$, can be perfectly estimated based on (\ref{eqn:solution 1}). Next, we intend to solve the equations given in (\ref{eqn:eta}). With known $\eta_{n,i}$'s, (\ref{eqn:eta}) characterizes a linear system with $(K-1)N$ variables and $N\tau_3$ equations. Therefore, a unique solution to (\ref{eqn:eta}) exists only when the number of equations is no smaller than the number of variables, i.e., $\tau_3\geq K-1$.

Next, we show that if $\tau_3=K-1$, there always exists a unique solution to (\ref{eqn:received pilot phase 3}) in the case of $M\geq N$. Specifically, if $\tau_3=K-1$, we can set $a_{k,i}$'s and $\phi_{n,i}$'s as given in (\ref{eqn:a case 1}) and (\ref{eqn:b case 1}), respectively. It then follows from (\ref{eqn:received pilot phase 3}) that
\begin{align}\label{eqn:solution 2}
\bar{\mv{y}}^{(\tau_1+\tau_2+k-1)}=[\mv{g}_{1,1},\cdots,\mv{g}_{1,N}]\mv{\lambda}_k, ~~~ k=2,\cdots,K.
\end{align}Since $\mv{g}_{1,n}$'s are linearly independent with each other with probability one in the case $M\geq N$, the pseudo-inverse matrix of $[\mv{g}_{1,1},\cdots,\mv{g}_{1,N}]$ exists. As a result, if $\tau_3=K-1$, there exists a unique solution to (\ref{eqn:received pilot phase 3}), which is given in (\ref{eqn:estimation 1}).

To summarize, in the case of $M\geq N$, there exists a unique solution to (\ref{eqn:received pilot phase 3}) only if $\tau_3 \geq K-1$. Moreover, $\tau_3 =K-1$ is sufficient to guarantee the existence of a unique solution to (\ref{eqn:received pilot phase 3}) by setting $a_{k,i}$'s and $\phi_{n,i}$'s according to (\ref{eqn:a case 1}) and (\ref{eqn:b case 1}). As a result, if $M\geq N$, $\tau_3=K-1$ is the minimum value of $\tau_3$ for perfectly estimating $\mv{\lambda}$ according to (\ref{eqn:received pilot phase 3}).

\subsection{Proof of Theorem \ref{theorem2}}\label{appendix2}
Similar to the case with $M\geq N$, we first prove that in the case of $M<N$, there exists a unique solution to (\ref{eqn:received pilot phase 3}) only if $\tau_3 \geq \lceil \frac{(K-1)N}{M}\rceil$. Note that in (\ref{eqn:received pilot phase 3}), the number of variables and the number of linear equations are $(K-1)N$ and $\tau_3 M$, respectively. As a result, there exists a unique solution to (\ref{eqn:received pilot phase 3}) only if the number of equations is no smaller than that of variables, i.e., $\tau_3 \geq \lceil\frac{(K-1)N}{M}\rceil$.

Next, we show that when $\tau_3 = \lceil\frac{(K-1)N}{M}\rceil$, there always exists a solution to (\ref{eqn:received pilot phase 3}) in the case of $M<N$. The basic idea is that in each time instant $\tau_1+\tau_2+i$ with $i\leq (K-1)\rho$, only one user $k\geq 2$ sends a non-zero pilot symbol for estimating $\lambda_{k,n}$'s with $n\in \Lambda_{k,1}$ without any interference from other users' pilot symbols, while in each time instant $\tau_1+\tau_2+i$ with $(K-1)\rho+1\leq i \leq \tilde{\tau}_3$, multiple users transmit non-zero pilot symbols simultaneously for estimating $\lambda_{k,n}$'s with $n\in \Lambda_{k,2}$'s by eliminating the interference caused by $\lambda_{k,n}$'s with $n\in \Lambda_{k,1}$.

Specifically, at time instant $\tau_1+\tau_2+i$ with $i\leq (K-1)\rho$, we schedule user $k=\lceil \frac{i}{\rho} \rceil+1$ to transmit a pilot symbol $1$, and each of the other users to transmits a pilot symbol $0$, as shown in (\ref{eqn:pilot TDMA 1}). Moreover, only $M$ IRS elements in the set of $\Omega_i$ are switched on and their reflecting coefficients are set to be $1$ as shown in (\ref{eqn:reflecting coefficient TDMA}). In this case, it can be shown that (\ref{eqn:received pilot phase 3}) reduces to
\begin{align}
\bar{\mv{y}}^{(\tau_1+\tau_2+i)} =&\sqrt{p}[\mv{g}_{1,\Omega_i(1)},\hspace{-2pt}\cdots\hspace{-2pt},\mv{g}_{1,\Omega_i(M)}]\nonumber \\ &[\lambda_{\lceil\frac{i}{\rho}\rceil+1,\Omega_i(1)},\cdots,\lambda_{\lceil\frac{i}{\rho}\rceil+1,\Omega_i(M)}]^T.
\end{align}Under the channel model of $\mv{r}_n$'s and $t_{k,n}$'s shown in (\ref{eqn:IRS to BS channel}) and (\ref{eqn:user to IRS channel}), in the case of $M<N$, any $M$ out of $N$ $\mv{g}_{1,n}$'s are linearly independent of each other with probability 1. Therefore, there exists a unique solution to the above equation, which is given by (\ref{eqn:perfect lambda case 1}).

Next, we estimate $\lambda_{k,n}$'s with $n\in \Lambda_{k,2}$. In time instant $\tau_1+\tau_2+i$ with $i \geq (K-1)\rho+1$, all the users in the set $\mathcal{K}_i$ will transmit pilot symbols 1, while each of the other users transmits a pilot symbol $0$, as shown in (\ref{eqn:pilot TDMA 2}). Moreover, all the $M_i\leq M$ IRS elements in the set $\mathcal{N}_i$ are switched on and their reflecting coefficients are set to be $1$ as shown in (\ref{eqn:reflecting coefficient TDMA 2}). In this case, the effective received signal at this time instant given in (\ref{eqn:received pilot reflected channel user k 2}) reduces to
\begin{align}\label{eqn:received signal 2}
\bar{\mv{y}}^{(\tau_1+\tau_2+i)}=\sum\limits_{k\in \mathcal{K}_i}\sum\limits_{n\in \mathcal{N}_i}\sqrt{p}\lambda_{k,n}\mv{g}_{1,n}.
\end{align}Further, for each user $k\in \mathcal{K}_i$, $\lambda_{k,n}$'s with $n\in \Lambda_{k,1}$ have already been perfectly estimated based on (\ref{eqn:perfect lambda case 1}). As a result, their interference can be canceled from (\ref{eqn:received signal 2}) to get $\tilde{\mv{y}}^{(\tau_1+\tau_2+i)}$ shown in (\ref{eqn:signal after interference cancellation}). Moreover, under our construction of $\Lambda_{k,1}$'s and $\Lambda_{k,2}$'s presented prior to Theorem \ref{theorem2}, for any two users $k1,k2\in \mathcal{K}_i$, we have $\Lambda_{k1,2}\cap\Lambda_{k2,2}=\emptyset$ and thus $\Lambda_{k1,2}\subset \Lambda_{k2,1}$ and $\Lambda_{k2,2}\subset \Lambda_{k1,1}$. It can then be shown that $\tilde{\mv{y}}^{(\tau_1+\tau_2+i)}$ reduces to
\begin{align}
\hspace{-5pt}\tilde{\mv{y}}^{(\tau_1+\tau_2+i)}=& \sqrt{p}[\mv{g}_{1,\mathcal{N}_i(1)},\cdots,\mv{g}_{1,\mathcal{N}_i(M_i)}]\nonumber \\ & \left[\hspace{-2pt}\lambda_{\lceil\frac{\mathcal{J}_i(1)}{\upsilon}\rceil\hspace{-1pt}+\hspace{-1pt}1,\mathcal{N}_i(1)},\cdots,\lambda_{\lceil\frac{\mathcal{J}_i(M)}{\upsilon}\rceil\hspace{-1pt}+\hspace{-1pt}1,\mathcal{N}_i(M)}\hspace{-2pt}\right]^T,
\end{align}where the $M_i$ elements in the set $\mathcal{N}_i$ can be shown to be different if $\Lambda_{k,2}$'s are constructed based on (\ref{eqn:Lambda 2}). As a result, there exists a unique solution to the above equation which is given by (\ref{eqn:perfect lambda case 2}).

To summarize, in the case of $M<N$, except for the last time instant, we are able to perfectly estimate $M$ unique $\lambda_{k,n}$'s either via (\ref{eqn:perfect lambda case 1}) or (\ref{eqn:perfect lambda case 2}) at each time instant, while at the last time instant, the remaining $\lambda_{k,n}$'s are estimated. As a result, the minimum $\tau_3$ for perfect channel estimation is characterized by (\ref{eqn:tau 3 case 2}). Theorem \ref{theorem2} is thus proved.

\end{appendix}

\bibliographystyle{IEEEtran}
\bibliography{IRS_channel_estimation}

\begin{IEEEbiography}[{\includegraphics[width=1in,height=1.25in,clip,keepaspectratio]{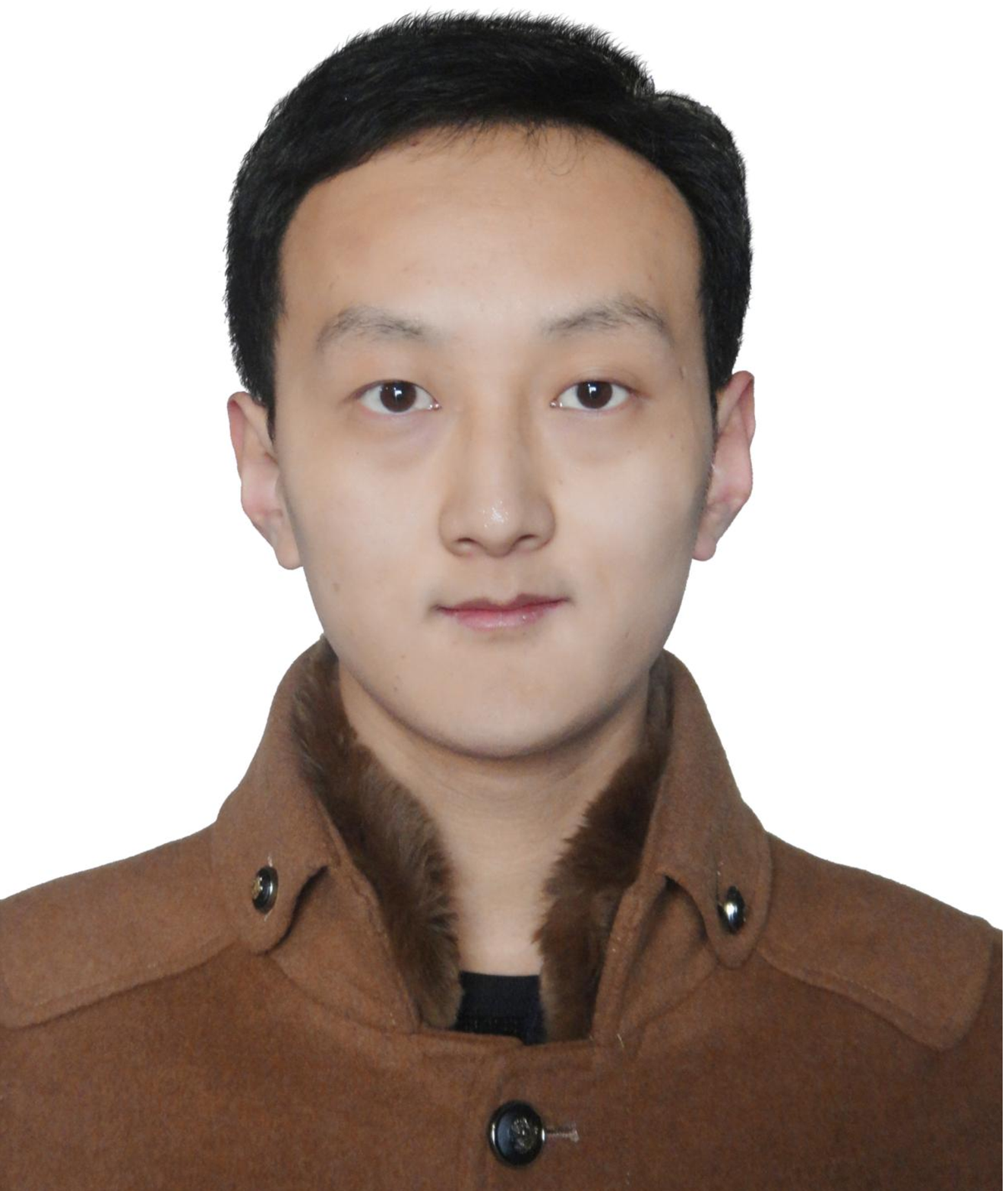}}] {Zhaorui Wang} (S'18-M'20) received his B.S. degree from University of Electronic Science and Technology of China in 2015, and the Ph.D. degree from The Chinese University of Hong Kong in 2019. He is now a postdoctoral fellow at The Hong Kong Polytechnic University. His research interests include intelligent reflecting surface (IRS) assisted communications, physical-layer network coding (PNC), and short packet communications. He was a recipient of the Hong Kong PhD Fellowship in 2015-2018.
\end{IEEEbiography}

\begin{IEEEbiography}[{\includegraphics[width=1in,height=1.25in,clip,keepaspectratio]{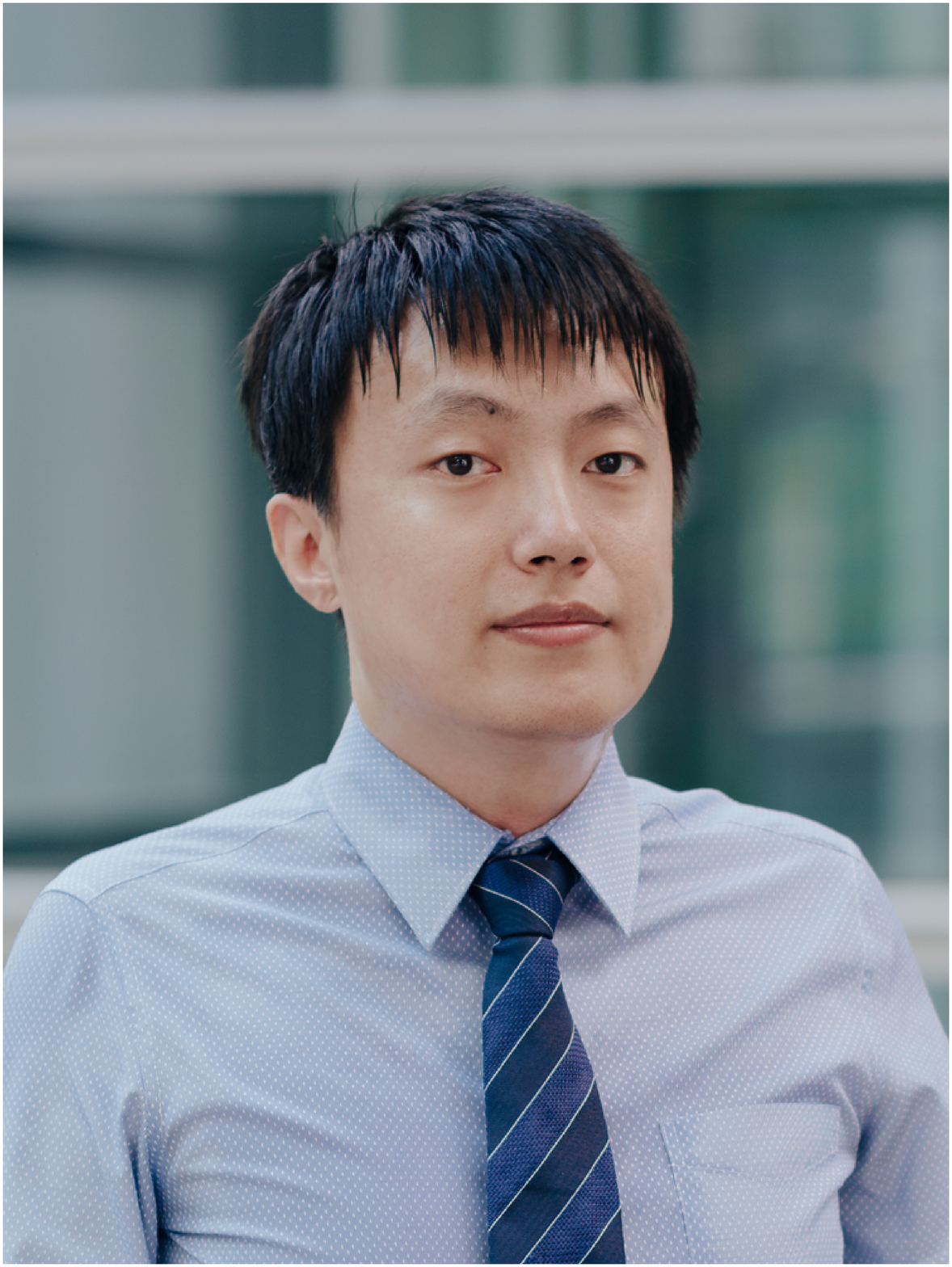}}] {Liang Liu} (S'14-M'15) received the B.Eng. degree from the Tianjin University, China, in 2010, and the Ph.D. degree from the National University of Singapore in 2014. He is currently an Assistant Professor in the Department of Electronic and Information Engineering at the Hong Kong Polytechnic University. Before that, he was a Research Fellow in the Department of Electrical and Computer Engineering at National University of Singapore from 2017 to 2018, and a Postdoctoral Fellow in the Department of Electrical and Computer Engineering at University of Toronto from 2015 to 2017.
	
His research interests include the next generation cellular technologies and machine-type communications for Internet of Things. He was the recipient of the IEEE Signal Processing Society Young Author Best Paper Award, 2017, and a Best Paper Award from IEEE WCSP in 2011. He is recognized by Clarivate Analytics as a Highly Cited Researcher, 2018.
\end{IEEEbiography}

\begin{IEEEbiography}[{\includegraphics[width=1in,height=1.25in,clip,keepaspectratio]{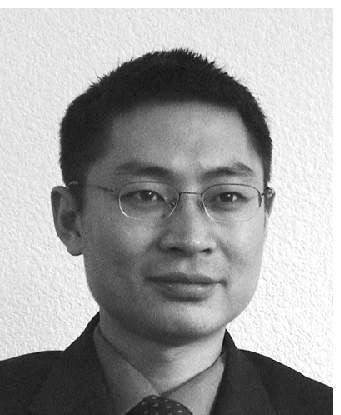}}] {Shuguang Cui} (S'99-M'05-SM'12-F'14) received his Ph.D in Electrical Engineering from Stanford University, California, USA, in 2005. Afterwards, he has been working as assistant, associate, full, Chair Professor in Electrical and Computer Engineering at the Univ. of Arizona, Texas A\&M University, UC Davis, and CUHK at Shenzhen respectively. He has also been the Executive Vice Director at Shenzhen Research Institute of Big Data. His current research interests focus on data driven large-scale system control and resource management, large data set analysis, IoT system design, energy harvesting based communication system design, and cognitive network optimization. He was selected as the Thomson Reuters Highly Cited Researcher and listed in the Worlds' Most Influential Scientific Minds by ScienceWatch in 2014. He was the recipient of the IEEE Signal Processing Society 2012 Best Paper Award. He has served as the general co-chair and TPC co-chairs for many IEEE conferences. He has also been serving as the area editor for IEEE Signal Processing Magazine, and associate editors for IEEE Transactions on Big Data, IEEE Transactions on Signal Processing, IEEE JSAC Series on Green Communications and Networking, and IEEE Transactions on Wireless Communications. He has been the elected member for IEEE Signal Processing Society SPCOM Technical Committee (2009-2014) and the elected Chair for IEEE ComSoc Wireless Technical Committee (2017-2018). He is a member of the Steering Committee for IEEE Transactions on Big Data and the Chair of the Steering Committee for IEEE Transactions on Cognitive Communications and Networking. He was also a member of the IEEE ComSoc Emerging Technology Committee. He was elected as an IEEE Fellow in 2013, an IEEE ComSoc Distinguished Lecturer in 2014, and IEEE VT Society Distinguished Lecturer in 2019.
\end{IEEEbiography}

\end{document}